\documentclass[twocolumn]{aastex631}
\usepackage{amsmath}
\usepackage{amssymb}
\def\code#1{\texttt{#1}}

\definecolor{purple}{rgb}{0.5,0,0.5}
\definecolor{darkgreen}{rgb}{0.1,0.6,0.1}
\definecolor{orange}{rgb}{1,0.6,0}

\newcommand{\brvs}{Br\"unt-V\"ais\"al\"a}

\submitjournal{ApJ}
\shorttitle{Skye EOS}
\shortauthors{Jermyn et al.}
\newcommand{\skye}{Skye}

\begin{document}

\title{Skye: A Differentiable Equation of State}

%% Note that the corresponding author command and emails has to come
%% before everything else. Also place all the emails in the \email
%% command instead of using multiple \email calls.
\correspondingauthor{Adam S. Jermyn}
\email{adamjermyn@gmail.com}

\author[0000-0001-5048-9973]{Adam S. Jermyn}
\affiliation{Center for Computational Astrophysics, Flatiron Institute, New York, NY 10010, USA}

\author[0000-0002-4870-8855]{Josiah Schwab}
\affiliation{Department of Astronomy and Astrophysics, University of California, Santa Cruz, CA 95064, USA}

\author[0000-0002-4791-6724]{Evan Bauer}
\affiliation{Center for Astrophysics $\vert$ Harvard \& Smithsonian, 60 Garden St Cambridge, MA 02138, USA}

\author[0000-0002-0474-159X]{F.X.~Timmes}
\affiliation{School of Earth and Space Exploration, Arizona State University, Tempe, AZ 85287, USA}
\affiliation{Joint Institute for Nuclear Astrophysics - Center for the Evolution of the Elements, USA}

\author[0000-0001-9955-4684]{Alexander Y. Potekhin}
\affiliation{Ioffe Institute, Politekhnicheskaya 26, Saint Petersburg 194021, Russia}

\begin{abstract}
Stellar evolution and numerical hydrodynamics simulations depend critically on access to fast, accurate, thermodynamically consistent equations of state.
We present \skye, a new equation of state for fully-ionized matter.
\skye\ includes the effects of positrons, relativity, electron degeneracy, Coulomb interactions, non-linear mixing effects, and quantum corrections.
\skye\ determines the point of Coulomb crystallization in a self-consistent manner, accounting for mixing and composition effects automatically.
A defining feature of this equation of state is that it uses analytic free energy terms and provides thermodynamic quantities using automatic differentiation machinery.
Because of this, \skye\ is easily extended to include new effects by simply writing new terms in the free energy.
We also introduce a novel \emph{thermodynamic extrapolation} scheme for extending analytic fits to the free energy beyond the range of the fitting data while preserving desirable properties like positive entropy and sound speed.
We demonstrate \skye\ in action in the \code{MESA} stellar evolution software instrument by computing white dwarf cooling curves.
\end{abstract}


\section{Introduction} \label{sec:intro}

\begin{figure*}
	\centering
	\includegraphics[width=0.9\textwidth]{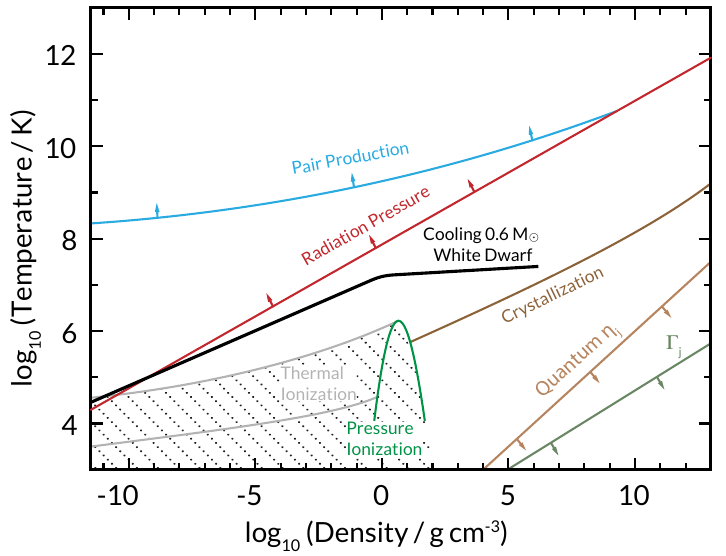}
	\caption{Coverage of the Skye EOS in the ($\rho,T$) plane. Shown is approximately where radiation pressure (red) 
                 dominates the gas pressure, thermodynamics from $e^-e^+$ pair production (light blue) dominates, crystallization
                 of ions (brown) begins, thermal (light gray) and pressure (green) ionization of atoms occurs.
%                 Regions where degeneracy (defined by $\mu_e/kT$, where $\mu_e$ is the electron chemical potential) 
%                 and special relativity (defined by $kT/m_ec^2$) are dominant are labeled (dark gray).
                 Lines of constant ion quantum parameter $\eta_j$ (light brown)
                 and ion interaction strength $\Gamma_j$ (dark green) are indicated in the lower-right, and attached arrows denote directions of increasing $\eta_j$ and $\Gamma_j$.
                 The dotted region marks where Skye's assumption of full ionization is a poor approximation.
                 An example profile, from core to surface, of a cooling white dwarf (black) is illustrated.}
	\label{fig:skye_eos}
\end{figure*}

The equation of state (EOS) of ionized matter is a key ingredient in models of stars, gas giant planets, accretion disks, and many other astrophysical systems.
These applications span many orders of magnitude in both density and temperature, and include both low-density systems that are thermally ionized (e.g., stellar atmospheres) and high-density ones that are pressure-ionized (e.g., planetary interiors).
Moreover matter can have many different compositions, ranging from pure hydrogen to exotic mixtures of heavy metals.
As a result, approximations to nature's EOS of ionized matter must capture a wide variety of physics (Figure~\ref{fig:skye_eos}) including relativity, quantum mechanics, electron degeneracy, pair production, phase transitions, and chemical mixtures.

Despite these challenges, several different equations of state have been introduced for ionized matter
\citep[e.g.,][]{1961ApJ...134..669S,eggleton_1973_aa,bludman_1977_aa,daeppen_1990_aa,pols_1995_aa,rogers_1996_aa,blinnikov_1996_aa,timmes_1999_aa,gong_2001_ab,dappen_2010_aa}.
\citet{1990JPhys..51.1607C} introduced an EOS for non-relativistic ionized hydrogen, incorporating sophisticated quantum and electron screening corrections.
Improvements then led to the PC EOS~\citep{CP1998,PhysRevE.62.8554,2009PhRvE..80d7401P,Potekhin2010}.
PC allows for arbitrary compositions and incorporates relativistic ideal electrons as well as modern prescriptions for electron screening and multi-component plasmas.
\citet{2013A&A...550A..43P} extended the PC EOS to include the effects of strong magnetic fields such as those found in neutron stars.
One of the distinguishing features of the PC EOS is the use of analytic prescriptions to capture non-ideal physics.

One of the limitations of the PC EOS is that it does not capture the effects of electron-positron pair production at high temperatures, which is important for the pair instability in massive stars~\citep{1967ApJ...148..803R}.
The treatment of electron degeneracy and the ideal quantum electron gas is also approximate, based on fitting formulas which approximate the relevant Fermi integrals.
These limitations are addressed by the HELM EOS~\citep{Timmes2000}.
While HELM does not include the sophisticated non-ideal corrections which are a defining strength of PC, it provides a tabulated Helmholtz free energy treatment of an ideal quantum electron-positron plasma, obtained by high-precision evaluation of the relevant Fermi-Dirac integrals \citep{cloutman_1989_aa,aparicio_1998_aa,gong_2001_aa}.
As such, HELM accurately and efficiently handles relativistic effects, degeneracy effects, and high-temperature pair production.

In this article we build on this progress by presenting a new equation of state, \skye, an EOS designed to handle density and temperature inputs over the range $10^{-12}\,\mathrm{g\,cm^{-3}} < \rho < 10^{13}\,\mathrm{g\,cm^{3}}$ and $10^3\,\mathrm{K} < T < 10^{13}\,\mathrm{K}$ (Figure~\ref{fig:skye_eos}).
\skye \ assumes material is fully-ionized, so the suitability
of the result is subject to the (composition-dependent) constraint that material is either pressure-ionized ($\rho \gtrsim 10^3\,\mathrm{g\,cm^{-3}}$) or thermally-ionized ($T \gtrsim 10^{5}\,\mathrm{K}$)\footnote{See Section~\ref{sec:limitations} for detailed composition-dependent ionization limits.}.
Further limits to Skye's suitability can arise due to violations of its other physics assumptions.
Building on HELM, we use the full ideal equation of state for electrons and positrons, accounting for degeneracy and relativity.
Ions are assumed to be a classical ideal gas.
We then add non-ideal classical and quantum corrections to account for electron-electron, electron-ion, and ion-ion interactions following a multi-component ion plasma prescription.
These corrections are generally similar to those used by the PC EOS, though we have used updated physics prescriptions in some instances~\citep[e.g., those of][]{2019MNRAS.488.5042B}.

Thermodynamic quantities in \skye\ are derived from a Helmholtz free energy to ensure thermodynamic consistency.
Automatic differentiation machinery allows extraction of arbitrary derivatives from an analytic Helmholtz free energy, allowing \skye\ to provide the high-order derivatives needed for stellar evolution calculations~\citep[e.g.,][]{Paxton2011}.
We further leverage this machinery to make the EOS easily extensible: adding new or refined physics to \skye\ is as easy as writing a formula for the additional Helmholtz free energy. The often painstaking and error-prone process of taking and programming analytic first, second, and even third derivatives of the Helmholtz free energy is eliminated. 
In this way \skye\ is a \emph{framework} for rapidly developing and prototyping new EOS physics as advances are made in numerical simulations and analytic calculations. We emphasize that \skye\ is not tied to a specific set of physics choices; \skye\ in 10 years is unlikely to be the same as \skye\ as described in this article.

In addition to being a single EOS which can be used at both high temperatures, like HELM, and high densities, like PC, \skye\ currently includes two significant physical improvements.
First, whereas PC fixes the location of Coulomb crystallization of the ions, \skye\ picks between the liquid and solid phase to minimize the Helmholtz free energy.
This enables a self-consistent treatment of the phase transition, albeit one currently without chemical phase separation, and means that the Helmholtz free energy is continuous across the transition.
Secondly, we introduce the technique of \emph{thermodynamic extrapolation}, 
which provides a principled way to extend Helmholtz free energy fitting formulas beyond their original range of applicability
and thus enables comparisons of the liquid and solid phase Helmholtz free energies.

This paper is structured as follows.
Important symbols are defined in Table~\ref{tab:symbols}.
In Section~\ref{sec:free} we explain the various terms which contribute to the Helmholtz free energy in \skye, as well as the new handling of phase transitions (Section~\ref{sec:non-ideal}) and thermodynamic extrapolation (Section~\ref{sec:extrapolate}).
Section~\ref{sec:thermo} shows how we extract thermodynamic quantities from the Helmholtz free energy.
We also introduce auxiliary quantities which allow stellar evolution software instruments to incorporate the latent heat of the Coulomb crystallization in a smooth manner.
Section~\ref{sec:limitations} discusses some of the current physics limitations of \skye,
which is principally that it does not extend to cases of partially ionized or neutral matter, or dense nuclear matter \citep{hempel_2012_aa}.
Section~\ref{sec:autodiff} introduces our automatic differentiation machinery.
In Section~\ref{sec:examples} we compare \skye\ to the PC and HELM equations of state and evaluate the quality of derivatives and thermodynamic consistency in \skye.
We also calculate white dwarf cooling tracks and demonstrate that \skye\ properly accounts for the latent heat of crystallization
(Section~\ref{sec:cooling}).
In Section~\ref{sec:speed} we demonstrate that \skye\ has comparable runtime performance to PC, making it viable for use in stellar evolution calculations.
\skye\ is open source and open-knowledge, and Section~\ref{sec:avail} describes options for obtaining and using \skye.
We conclude with a discussion of future work in Section~\ref{sec:conc}.

\startlongtable
\begin{deluxetable}{llr}
  \tablecolumns{3}
  \tablewidth{0.5\textwidth}
  \tablecaption{Important symbols.\label{tab:symbols}}
  \tablehead{\colhead{Name} & \colhead{Description} & \colhead{Appears}}
  \startdata
$T$ & Temperature & \ref{sec:intro}\\
$\rho$ & Density & \ref{sec:intro}\\
$F$ & Helmholtz Free Energy & \ref{sec:free}\\
$F_{\rm ideal}$ & Ideal Free Energy & \ref{sec:free}\\
$F_{\textrm{non-ideal}}$ & Non-ideal Free Energy & \ref{sec:free}\\
$F_{\rm rad}$ & Radiation Gas Free Energy & \ref{sec:ideal}\\
$F_{\rm ideal \ e^- e^+}$ & Ideal Electron-Positron Free Energy & \ref{sec:ideal}\\
$F_{\rm ideal \ ion}$ & Ideal Ion Free Energy & \ref{sec:ideal}\\
$F_{\rm ideal \ mix}$ & Ideal Ion Mixing Free Energy & \ref{sec:ideal}\\
$a$ & Radiation Gas Constant & \ref{sec:ideal}\\
$k_{\rm B}$ & Boltzmann Constant & \ref{sec:ideal}\\
$m_j$ & Mass of species $j$ & \ref{sec:ideal}\\
$y_j$ & Number fraction of ion species $j$ & \ref{sec:ideal}\\
$\bar{m}$ & Average ion mass & \ref{sec:ideal}\\
$n_j$ & Number density of species $j$ & \ref{sec:ideal}\\
$n_{Q,j}$ & Quantum density of ion species $j$ & \ref{sec:ideal}\\
$M_{\rm spin}$ & Spin multiplicity of ion species $j$ & \ref{sec:ideal}\\
$\hbar$ & Reduced Planck Constant & \ref{sec:ideal}\\
$a_j$ & Sphere radius of species $j$ & \ref{sec:non-ideal}\\
        & $(3n_j/4\pi)^{-1/3}$&\\
$r_{s,j}$ & Non-dimensional radius of species $j$ & \ref{sec:non-ideal}\\
        &$Z_j^2 m_j e_j^2 a_j / \hbar^2$&\\
$Z_j$ & Charge of species $j$  & \ref{sec:non-ideal}\\
        &($-1$ for electrons)&\\
$\Gamma_j$ & Coupling parameter of species $j$ & \ref{sec:non-ideal}\\
        &$Z_j^2 e^2 / a_j k_{\rm B} T$&\\
$\eta_j$ & Quantum Parameter of species $j$ & \ref{sec:non-ideal}\\
        &$(\hbar/k_{\rm B} T)\sqrt{4\pi e^2 n_j Z_j^2/m_j}$&\\
$p_{\rm F}$ & Fermi Momentum & \ref{sec:non-ideal}\\
$x_r$ & Relativity Parameter $p_{\rm F}/m_e c$ & \ref{sec:non-ideal}\\
$\gamma$ & Fermi Lorentz Factor & \ref{sec:non-ideal}\\
        & $\sqrt{1+x_r^2}$&\\
$E_{\rm F}^{\rm NR}$ & Non-relativistic Fermi Energy & \ref{sec:non-ideal}\\
$h(\alpha)$ & Switch function & \ref{sec:non-ideal}\\
$\alpha$ & Switch parameter & \ref{sec:non-ideal}\\
        &$3 k_{\rm B} T \gamma / 2 E_{\rm F}^{\rm NR}$&\\
$e$ & Specific internal energy & \ref{sec:extrapolate}\\
$s$ & Specific entropy & \ref{sec:extrapolate}\\
$T_{\rm b}$ & Extrapolation Temperature & \ref{sec:extrapolate}\\
$\Gamma_{\rm max}^{\rm liquid}$ & Liquid extrapolation $\Gamma_j$& \ref{sec:extrapolate}\\
$\Gamma_{\rm min}^{\rm solid}$ & Solid extrapolation $\Gamma_j$& \ref{sec:extrapolate}\\
$p$ & Pressure & \ref{sec:thermo}\\
$c_v$ & Specific heat at constant volume & \ref{sec:thermo}\\
$c_p$ & Specific heat at constant pressure & \ref{sec:thermo}\\
$\chi_T$ & Thermal susceptibility & \ref{sec:thermo}\\
$\chi_\rho$ & Density susceptibility & \ref{sec:thermo}\\
$\Gamma_1$ & First adiabatic exponent & \ref{sec:thermo}\\
$\Gamma_2$ & Second adiabatic exponent & \ref{sec:thermo}\\
$\Gamma_3$ & Third adiabatic exponent & \ref{sec:thermo}\\
$\nabla_{\rm ad}$ & Adiabatic Gradient & \ref{sec:thermo}\\
$c_s$ & Sound speed & \ref{sec:thermo}\\
$\phi$ & Smoothed phase parameter & \ref{sec:thermo}\\
$L_T$ & Latent $T ds/d\ln T$ & \ref{sec:thermo}\\
$L_\rho$ & Latent $T ds/d\ln \rho$ & \ref{sec:thermo}\\
$T_{j}^{\rm ion}$ & Full-ionization $T$ of species $j$ & \ref{sec:limitations}\\
$\rho_{j}^{\rm ion}$ & Full-ionization $\rho$ of species $j$ & \ref{sec:limitations}\\
$\rho_{j,\rm nuclear}$ & Nuclear density of species $j$ & \ref{sec:limitations}\\
$T_{\rm QCD}$ & Temperature of & \ref{sec:limitations}\\
 & proton rest mass-energy & \\
  \enddata
\end{deluxetable}

\section{Helmholtz Free Energy} \label{sec:free}

The \skye\ equation of state is based on a Helmholtz free energy $F(\rho,T,\{n_j\})$ given by
\begin{align}
	F = F_{\rm ideal} + F_{\textrm{non-ideal}},
\end{align}
where $n_j$ is the number density of species $j$.
Here $F$ is in terms of energy per unit mass.
The ideal term incorporates all non-interacting contributions of relativistic electrons and positrons, non-relativistic non-degenerate ions, and photons.
The non-ideal term contains the contributions of Coulomb interactions among and between electrons and ions.

\subsection{Ideal Terms}\label{sec:ideal}

The ideal free energy is
\begin{align}
	F_{\rm ideal} = F_{\rm rad} + F_{\rm ideal \ e^- e^+} + F_{\rm ideal\ ion} + F_{\rm ideal\ mix}.
\end{align}

$F_{\rm rad}$ is the free energy of an ideal gas of photons, 
\begin{align}
	F_{\rm rad} = -\frac{a T^4}{3 \rho},
\end{align}
where $a$ is the radiation gas constant.

$F_{\rm ideal\ e^- e^+}$ represents an ideal gas of non-interacting electrons and positrons, 
obtained from biquintic Hermite polynomial interpolation of a table (\citealt{Timmes2000}, also see \citealt{baturin_2019_aa}).
This single table captures both relativistic and degeneracy effects and is valid for any 
fully ionized composition.

$F_{\rm ideal\ ion}$ represents an ideal gas of non-degenerate ions and is given by~\citep[see e.g.][]{Potekhin2010}
\begin{align}
	F_{\rm ideal\ ion} = \frac{k_{\rm B} T}{\bar{m}} \sum_j y_j \left[\ln \left(\frac{n_j}{n_{Q,j}}\right) - 1\right],
	\label{eq:Fidealion}
\end{align}
where $y_j$ is the number fraction of species $j$,
\begin{align}
	\bar{m} \equiv \sum_y y_j m_j
\end{align}
is the mean ionic mass in $\mathrm{g}$, $m_j$ is the mass of ion species $j$, and
\begin{align}
	n_{Q,j} \equiv M_{\rm spin,j}\left(\frac{2\pi \hbar^2}{m_j k_{\rm B} T}\right)^{-3/2}.
\end{align}
Here $M_{\rm spin,j}$ is the spin multiplicity of the ion.
The effect of $M_{\rm spin,j}$ is to introduce a composition-dependent offset in the entropy and so for simplicity we neglect it, setting $M_{\rm spin,j}=1$.

$F_{\rm ideal\ mix}$ captures the ideal free energy of mixing for ions, given by
\begin{align}
	F_{\rm ideal\ mix} = \frac{k_{\rm B} T}{\bar{m}} \sum_j y_j \ln y_j.
\end{align}

\subsection{Non-Ideal Terms}\label{sec:non-ideal}

The non-ideal free energy of electron interactions is commonly written in terms of the electron interaction strength
\begin{align}
	\Gamma_e \equiv \frac{e^2}{a_e k_{\rm B} T},
\end{align}
where
\begin{align}
	a_e \equiv \left(\frac{4}{3} \pi n_e\right)^{-1/3},
\end{align}
and $n_e$ is the electron number density.
Likewise the ion interaction free energy is given in terms of the ion interaction strength
\begin{align}
	\Gamma_j \equiv \Gamma_e Z_j^{5/3},
\end{align}
where $Z_j$ is the charge of ion species $j$.
The average Coulomb parameter is
\begin{align}
	\langle \Gamma \rangle = \sum_j y_j \Gamma_j .
\end{align}
Finally, quantum effects enter for ions via the parameter
\begin{align}
	\eta_j \equiv \frac{T_{{\rm p},j}}{T} = \frac{\hbar}{k_{\rm B}T}\sqrt{\frac{4\pi e^2 n_j Z_j^2}{m_j}},
	\label{eq:eta}
\end{align}
which is proportional to $\Gamma_j \lambda / a_j$, where $\lambda$ is the De-Broglie wavelength of a non-relativistic particle.
In these terms we write
\begin{align}
	F_{\textrm{non-ideal}} = \frac{k_{\rm B} T}{\bar{m}}&\left[f_{\rm e-e}(\Gamma_e, \eta_e) \right.\\
	& \left. + f_{\rm i}(\{Z_j\}, \{m_j\}, \{\Gamma_j\},\{\eta_j\})\right]\nonumber,
\end{align}	
where each $f$ is a free energy per ion per $k_{\rm B} T$ and $\eta_e$ is the electronic quantum parameter, given by using the electron mass and $Z_e=1$ in equation~\eqref{eq:eta}.
While the symbol $\eta$ or $\eta_e$ is also commonly used to represent the electron degeneracy, we never do so in this paper.

$f_{\rm e-e}$ is the free energy of Coulomb interactions between electrons, also known as the electron-exchange energy.
We implement this via the non-relativistic formula of~\citet{ICHIMARU198791}, which~\citet{Potekhin2010} argued should suffice because in highly relativistic scenarios the electron-exchange energy is a small part of the total.

$f_{\rm i}$ captures non-ideal effects associated with mixing, Coulomb interaction among ions, and Coulomb interactions between ions and electrons (i.e., polarization or screening).
Because an interacting Coulomb gas can crystallize, we compute this term twice, once assuming the liquid phase and once assuming the solid phase.
We then take
\begin{align}
	f_{\rm i} = \min(f_{\rm i}^{\rm liquid}, f_{\rm i}^{\rm solid}),
	\label{eq:fmin}
\end{align}
so as to minimize the free energy across the possible options.%
\footnote{In stars, the phase transition technically occurs at constant pressure rather than
  constant volume and so minimizes the Gibbs free energy.
  Appendix~A in \citet{2010PhRvE..81c6107M} demonstrates that minimizing the Helmholtz free energy instead
  does not significantly affect the phase diagram.}

\subsubsection{Liquid Phase}

In the liquid phase we decompose $f_{\rm i}$ as
\begin{align}
	f_{\rm i}^{\rm liquid} = f_{\rm mix}^{\rm liquid} + \sum_j y_j (f_{\rm OCP, j}^{\rm classical} + f_{\rm OCP, j}^{\rm quantum} + f_{\rm i-e, j}^{\rm liquid}),
	\label{eq:liq_f}
\end{align}
where $f_{\rm mix}^{\rm liquid}$ captures non-ideal corrections to the mixing free energy in the liquid phase, the $f_{\rm OCP,j}$ terms represent the free energy of a one-component plasma (OCP) made entirely of species $j$, and $f_{\rm i-e,j}$ accuonts for electron-ion interactions for species $j$.

We obtain $f_{\rm OCP, j}^{\rm classical}$ from the fit of~\citet{PhysRevE.62.8554} with the parameter set matching the Monte Carlo calculations of~\citet{1999CoPP...39...97D}, which were performed over $1 \leq \Gamma_j \leq 200$.
This fit matches the Debye-H\"{u}ckel approximation at low $\Gamma_j$ as well as leading-order corrections to this approximation, so these fits are valid for $\Gamma_j \leq 200$.

We chose this particular classical fit because it is the same one~\citet{2019MNRAS.490.5839B} used to derive the quantum correction $f_{\rm OCP, j}^{\rm quantum}$, which was fit to path-integral Monte Carlo calculations performed over $1 \leq \Gamma_j \leq 175$ and $600 \leq r_{s,j} \leq 120,000$~\citep{2019MNRAS.488.5042B}, where
\begin{align}
	r_{s,j} \equiv \frac{m_j Z_j^2 e^2}{\hbar^2}\left(\frac{4}{3} \pi n_j\right)^{-1/3}
\end{align}
is the dimensionless ion sphere radius.

We obtain $f_{\rm i-e,j}^{\rm liquid}$ using the formula of~\citet{PhysRevE.62.8554}, which was chosen to fit Hypernetted Chain calculations on the range $0 < \Gamma \la 300$ and $0 < r_{s,e} < 1$, where
\begin{align}
	r_{s,e} \equiv \frac{m_j e^2}{\hbar^2}\left(\frac{4}{3} \pi n_e\right)^{-1/3}
\end{align}
is the dimensionless electron sphere radius.

\citet{2009PhRvE..80d7401P} computed classical corrections to the linear mixing rule using Hypernetted Chain calculations.
These were combined with the Monte Carlo calculations of~\citet{1999JChPh.111.9695C} to produce a data set spanning $10^{-3} < \Gamma_j < 10^2$.
\citet{2009PhRvE..80d7401P} then produced an analytic fitting formula matching these data.
The form was chosen to reproduce analytic expectations in the limits of both large and small $\Gamma_j$.
We use this fit for $f_{\rm mix}^{\rm liquid}$.

\subsubsection{Solid Phase}

In the solid phase we use a similar decomposition:
\begin{align}
	f_{\rm i}^{\rm solid} = f_{\rm mix}^{\rm solid} + \sum_j y_j (f_{\rm OCP, j}^{\rm harmonic} + f_{\rm OCP, j}^{\rm anharmonic} + f_{\rm i-e,j}^{\rm solid}),
	\label{eq:sol_f}
\end{align}
where $f_{\rm mix}^{\rm solid}$ captures non-ideal corrections to the mixing free energy in the solid phase and is formed by summing contributions pairwise between species, $f_{\rm OCP, j}^{\rm harmonic}$ represents the harmonic crystal free energy (i.e., phonons), $f_{\rm OCP, j}^{\rm anharmonic}$ captures anharmonic corrections, and $f_{\rm i-e,j}^{\rm solid}$ provides the free energy of electron-ion interactions (i.e., screening/polarization).

The harmonic free energy is given by calculations due to~\citet{PhysRevE.64.057402} and is valid at any $\Gamma_j$ where the system takes on a crystal structure.
Because the body centered cubic (BCC) lattice has the lowest free energy of the ones they consider we use their BCC coefficients.

The anharmonic free energy is given by a sum of a classical term from~\citet{PhysRevE.47.4330} and quantum corrections from~\citet{Potekhin2010}.
The classical term is an analytic fit to Monte Carlo data over the range $170 \leq \Gamma_j \leq 2000$, and the form of the fit was chosen to match expectations from perturbation theory in the large-$\Gamma_j$ limit, so this term should be valid for $\Gamma_j \geq 170$.
The quantum corrections are a combination of terms meant to reproduce analytic expansions about the classical~\citep[$\eta_j\rightarrow 0$]{HANSEN1975187} and zero-temperature~\citep[$\Gamma_j/\sqrt{\eta_j} \rightarrow \infty$]{PhysRevA.36.1859,1961PhRv..124..747C} limits.
At fixed $\Gamma_j$ these are opposing limits in $\eta_j$, so in principle these corrections may be used at any $\eta_j$.

For the solid mixing free energy we support the formulas of either~\citet{1993PhRvE..48.1344O} or~\citet{2013A&A...550A..43P}, extended from the three-component case to many component plasmas following~\citet{2010PhRvE..81c6107M}.
The formula of~\citet{1993PhRvE..48.1344O} was produced to match Monte Carlo calculations of crystals performed at charge ratios $4/3 \leq R \leq 4$, where $R$ is the ratio of the charge of the higher-$Z$ species to that of the lower-$Z$ one, while that of~\citet{2013A&A...550A..43P} was designed to match both the results of~\citet{1993PhRvE..48.1344O} and~\citet{2003CoPP...43..279D}.
In either case the fit is linear in $\Gamma$ because only the Madelung energy is considered in the Monte Carlo calculations, and this is linear in $\Gamma$ by construction.
We apply this formula by grouping all species of a given charge together, because the scheme of~\citet{2010PhRvE..81c6107M} is independent of species mass and just captures corrections to the potential energy of a multicomponent plasma.

We obtain $f_{\rm i-e,j}^{\rm solid}$ using the formula of~\citet{Potekhin2010}, which was fitted to numerical calculations by~\citet{PhysRevE.62.8554} on the range $80 < \Gamma \la 3\times 10^4$ and $10^{-2} < x_r < 10^2$, where $x_r$ is the relativity parameter
\begin{align}
	x_r \equiv \frac{p_{\rm F}}{m_e c}
\end{align}
for Fermi momentum $p_{\rm F}$, electron mass $m_e$, and speed of light $c$.
This formula is based on a perturbation expansion which is known to break down at low densities~\citep{1976PhRvA..14..816G}.
In particular, the expression for $f_{\rm i-e,j}^{\rm solid}$ in the solid phase was tested up to $x_r \ga 10^{-2}$, corresponding to densities of $\rho \ga 1\,\mathrm{g\,cm^{-3}} (m_j / Z_j m_p)$.
Unlike the liquid phase formula, however, this one does not reproduce the Debye-H\"{u}ckel limit at low densities, and rises without bound like $\rho^{-1/3}$ towards low densities.
Moreover it diverges at low $\Gamma$ and so cannot be used for $\Gamma \la 80$.

To remedy this we smoothly transition from the solid screening formula to the liquid screening formula, which reproduces the appropriate high-temperature and low-density limits.
We do this by writing
\begin{align}
	f_{\rm i-e,j}^{\rm solid} =  h(\alpha) f_{\rm i-e,j}^{\rm liquid} + (1-h(\alpha)) f_{\rm i-e,j}^{\rm solid, original},
\end{align}
where
\begin{align}
	h(\alpha) \equiv \tanh^3(2\alpha)
\end{align}
is a smooth switch function and
\begin{align}
	\alpha \equiv \frac{3 k_{\rm B} T \gamma}{2 E_{\rm F}^{\rm NR}} = 3\left(\frac{4}{9\pi}\right)^{2/3} \frac{r_s}{\Gamma_e} \gamma
\end{align}
measures the degeneracy of the system, becoming large in the Debye-H\"{u}ckel limit and small in the Thomas-Fermi limit.
Here $E_{\rm F}^{\rm NR}$ is the non-relativistic Fermi energy and $\gamma = \sqrt{1 + x_r^2}$ is the Lorentz parameter at the Fermi momentum.
We choose $\alpha$ for our switch because it controls whether the dielectric function more closely resembles the Debye-H\"{u}ckel or Thomas-Fermi limits.

\subsection{Thermodynamic Extrapolation}\label{sec:extrapolate}

In order to implement equation~\eqref{eq:fmin} we need to be able to evaluate all components of the free energy at any point in the $(\rho,T)$ plane.
Unfortunately, the fits we use for the one-component plasma $f_{\rm OCP}$ have limited ranges of validity.
For instance the classical liquid free energy was fit to Monte Carlo simulations in the range $1 \leq \Gamma_j \leq 200$.
The low-$\Gamma_j$ asymptotic behavior is known analytically and enforced by the fitting formula, but the high-$\Gamma_j$ behavior ($\Gamma_j > 200$) is in a sense undefined: beyond crystallization it is not obvious what it means to speak of a liquid free energy.
The same is true of the solid phase free energy formula, which was computed via a perturbation expansion in $1/\Gamma_j$ and diverges at small $\Gamma_j$.

This problem is not just mathematical, it is conceptual: any scheme which extends these formulas beyond their range of validity makes implicit assumptions about the physical behavior of the system, and there is no guarantee that following the analytic behavior of the fitting formulas will happen to capture the right physics.
Indeed, as mentioned, many of these fitting formulae diverge away from the limits for which they were designed.

To address this we make our choice of physics explicit.
For the liquid phase free energy we assume that the probability distribution over microscopic states is fixed for $\Gamma_j > \Gamma_{\rm max}^{\rm liquid}=200$.
For the solid phase free energy we make the same assumption when $\Gamma_j < \Gamma_{\rm min}^{\rm solid}=170$.
This assumption amounts to an ansatz: we \emph{define} a high-$\Gamma_j$ liquid to be characterized by the probability distribution of $\Gamma_{\rm max}^{\rm liquid}$, and likewise for a low-$\Gamma_j$ solid with $\Gamma_{\rm min}^{\rm solid}$.
These ranges were chosen to permit using the OCP terms over the widest range over which each free energy component in equations~\eqref{eq:liq_f} and~\eqref{eq:sol_f} are known to be accurate.

Because the energy is given by the ensemble average
\begin{align}
	e(\Gamma,\eta) = \sum_s p_s(\Gamma_j,\eta_j) e_s,
\end{align}
where $p_s$ and $e_s$ are the probability and energy of microstate $s$, an immediate consequence of our choice to fix $p_s$ out-of-bounds is that the energy must be constant.
Similarly the specific entropy
\begin{align}
	s = -\frac{k_{\rm B} T}{\bar{m}}\sum_s p_s \ln p_s
\end{align}
is constant out-of-bounds because $p_s$ is fixed.

That is,
\begin{align}
	\left.\frac{\partial s}{\partial T}\right|_{\rho} = -\frac{\partial^2 F}{\partial T^2} = 0.
\end{align}
This condition combined with continuity of entropy and free energy at the boundary allows us to uniquely define an extrapolated free energy
\begin{align}
	F_{\rm ext.}(\rho,T) = F(\rho,T_{\rm b}(\rho)) + (T_{\rm b}(\rho)-T) s_{\rm b}(\rho),
\end{align}
where the subscript ``b'' denotes a quantity evaluated at the boundary.
Note that by construction this form also enforces $\partial e/\partial T=0$ out-of-bounds.

This prescription provides a robust extrapolation far beyond the limits of the original fitting formulas which avoids common extrapolation pitfalls such as negative entropies or sound speeds.
However, because $\partial s/\partial T$ and $\partial e/\partial T$ are forced to zero, this extrapolation scheme does produce discontinuities in quantities like the heat capacity.
We encounter these discontinuities in Section~\ref{sec:cooling} and, while they do not cause a problem there, in some applications it may be desirable to continue to apply the original fitting formulas slightly beyond the data on which they were based.

We currently apply this extrapolation scheme just to the classical and quantum ion-ion OCP terms and \emph{not} to the mixing corrections $f_{\rm mix}^{\rm liquid}$ and $f_{\rm mix}^{\rm solid}$ or to the electron-ion screening terms $f_{\rm i-e,j}^{\rm solid}$ and $f_{\rm i-e,j}^{\rm liquid}$.
The liquid mixing corrections are constructed to match analytic expectations in the limits of both large and small $\Gamma_j$, and the solid mixing corrections are linear in $\Gamma_j$ by construction because they only consider the Madelung energy.
As a result neither mixing correction requires extrapolation in $\Gamma_j$.
Likewise both sets of screening corrections obey the correct asymptotic limits at both large and small $\Gamma_j$ and so neither requires extrapolation.

Note that while this extrapolation scheme ensures that the relevant free energy terms are well-behaved in $\Gamma_j$, they may still exhibit unphysical asymptotic behaviour in $\eta_j$, i.e. towards very large or small densities.
This may be the cause of some of the unusual features we see in the phase diagram in Appendix~\ref{sec:quantum}.

\section{Thermodynamics} \label{sec:thermo}

\skye\ computes thermodynamic quantities from derivatives of the free energy $F=e-Ts$.
The entropy, pressure, and internal energy are given by
\begin{align}
	s &= -\left.\frac{\partial F}{\partial T}\right|_\rho\\
	e &= F + T s\\
	p &= \rho^2 \left.\frac{\partial F}{\partial \rho}\right|_T.
\end{align}
From the internal energy we obtain the specific heat at constant volume
\begin{align}
	c_v &= \left.\frac{\partial e}{\partial T}\right|_{\rho}
\end{align}
From the pressure we find the susceptibilities
\begin{align}
	\chi_T &\equiv \left.\frac{\partial \ln p}{\partial \ln T}\right|_{\rho}\\
	\chi_\rho &\equiv \left.\frac{\partial \ln p}{\partial \ln \rho}\right|_{T},
\end{align}
which then form the adiabatic indices and gradient~\citep{1968pss..book.....C}
\begin{align}
	\Gamma_3 \equiv 1 + \frac{p}{\rho c_v T} \chi_T\\
	\Gamma_1 \equiv \chi_\rho + (\Gamma_3 - 1)\chi_T\\
	\nabla_{\rm ad} \equiv \frac{\Gamma_3 - 1}{\Gamma_1}\\
	\Gamma_2 = 1 - \nabla_{\rm ad}.
\end{align}
Note that $\Gamma_{1,2,3}$ are \emph{not} ion interaction parameters but rather adiabatic indices.
From these we find the specific heat at constant pressure
\begin{align}
	c_p &= c_v \frac{\Gamma_1}{\chi_\rho}
\end{align}	
and the sound speed accounting for relativity~\citep{1968pss..book.....C}
\begin{align}
	c_s &= c \sqrt{\frac{\Gamma_1}{1 + \frac{\rho}{p}(e + c^2)}},
\end{align}
where $c$ is the speed of light.

\skye\ further reports several auxiliary quantities meant to help with calculations which cross the liquid-solid phase boundary.
Derivatives of the free energy may be discontinuous across the phase transition, which means that $s$, $e$, and $p$ may be discontinuous there.
This is a particular problem for stellar evolution calculations.

To understand the problem consider the term
\begin{align}
	\epsilon_{\rm grav} \equiv -T \frac{ds}{dt},
\end{align}
which commonly appears in the energy or heat equation in stellar evolution software instruments.
Here $d/dt$ denotes a Lagrangian derivative.
If $ds/dt$ is evaluated by finite differences then no time step will be small enough to produce a converged result across the phase transition because $s$ is genuinely discontinuous there.

On the other hand, if we write
\begin{align}
	\frac{ds}{dt} = \left.\frac{\partial s}{\partial T}\right|_\rho \frac{dT}{dt} + \left.\frac{\partial s}{\partial \rho}\right|_T \frac{d\rho}{dt},
\end{align}
then we miss the latent heat of the phase transition because, except for a set in $(\rho,T)$ of measure zero, $\partial s/\partial T$ and $\partial s/\partial \rho$ contain no information about the transition.
This is not a mathematical problem: near the phase transition $\partial s/\partial T \propto \delta(T-T_{\rm transition})$, and likewise for $\partial s/\partial \rho$.
The problem is that we cannot directly implement a Dirac delta function in numerical calculations, and neglecting this term means neglecting the latent heat of the transition.

To address this, in addition to equation~\eqref{eq:fmin} we also compute a smoothed version of the free energy
\begin{align}
	f_{\rm i, smooth} =  \phi f_{\rm i, liquid}+ (1-\phi) f_{\rm i, solid},
	\label{eq:phase_mix}
\end{align}
where
\begin{align}
	\phi = \frac{e^{\Delta f/w}}{e^{\Delta f/w} + 1}
	\label{eq:phi}
\end{align}
measures which phase the system is in, and smoothly transitions from the liquid phase to the solid phase across the crystallization boundary.
Here $w$ is a blurring parameter, which we choose to be $10^{-2}$ to ensure a narrow transition, and
\begin{align}
	\Delta f = f_{\rm i,liquid} - f_{\rm i,solid}.
\end{align}

The delta functions which appear in derivatives of $f_{\rm i}$ appear as smooth functions with broad support in $f_{\rm i, smooth}$.
Unfortunately this smoothed free energy also produces unphysical properties, such as negative sound speeds and entropies.
So we cannot use thermodynamic quantities derived from $f_{\rm i,smooth}$ directly in place of those derived from $f_{\rm i}$.
However, we can use $f_{\rm i,smooth}$ to calculate an additional heating term which compensates for the missing latent heat.

To see this let $T_s$ be the temperature where $\phi=\epsilon \ll 1$, let $T_t$ be the temperature where $\phi=1/2$, and let $T_l$ be the temperature where $\phi = 1 - \epsilon$.
The entropy difference between $T_s$ and $T_t$ is similar for both $s$ and $s_{\rm smooth}$, i.e.
\begin{align}
	s_{\rm smooth}(T_t) - s_{\rm smooth}(T_s) &\approx s(T_t) - s(T_s) + \mathcal{O}(\epsilon).
\end{align}
We can rewrite this in the form
\begin{align}
	\int_{T_s}^{T_l} \left.\frac{\partial s_{\rm smooth}}{\partial T}\right|_{\rho} - \left.\frac{\partial s_{\rm regular}}{\partial T}\right|_{\rho} - \Delta s \delta(T-T_{t}) dT &\approx \mathcal{O}\left(\epsilon\right),
\end{align}
where here the subscript ``regular'' means the part of the derivative excluding the Dirac delta, which we have included explicitly in the third term.
Rearranging this we find
\begin{align}
	\Delta s \approx \int_{T_s}^{T_l} \left.\frac{\partial s_{\rm smooth}}{\partial T}\right|_{\rho} - \left.\frac{\partial s_{\rm regular}}{\partial T}\right|_{\rho} dT +\mathcal{O}\left(\epsilon\right).
\end{align}

Using this formalism, we can write the latent heat which ought to appear in $\epsilon_{\rm grav}$ but which we would otherwise miss as
\begin{align}
	\epsilon_{\rm latent} &= T \left(\left.\frac{\partial s_{\rm smooth}}{\partial T}\right|_\rho-\left.\frac{\partial s_{\rm regular}}{\partial T}\right|_\rho\right) \frac{dT}{dt} \nonumber\\
	&+ T\left(\left.\frac{\partial s_{\rm smooth}}{\partial \rho}\right|_T-\left.\frac{\partial s_{\rm regular}}{\partial \rho}\right|_T\right) \frac{d\rho}{dt}
	\label{eq:Ls}
\end{align}
where $s_{\rm smooth}$ is the entropy calculated from the smoothed free energy.
To facilitate calculating $\epsilon_{\rm latent}$, \skye\ reports
\begin{align}
	L_T &\equiv T \left(\left.\frac{\partial s_{\rm smooth}}{\partial \ln T}\right|_\rho-\left.\frac{\partial s_{\rm regular}}{\partial \ln T}\right|_\rho\right)\\
	L_\rho &\equiv T \left(\left.\frac{\partial s_{\rm smooth}}{\partial \ln \rho}\right|_T-\left.\frac{\partial s_{\rm regular}}{\partial\ln  \rho}\right|_T\right),
\end{align}	
as well as the smoothed phase $\phi$ for diagnostic purposes.

\section{Limitations} \label{sec:limitations}

The physics in \skye\ models a fully-ionized multi-component quantum ion plasma, quantum and relativistic ideal electrons with non-ideal electron-electron interactions, and ideal radiation.
These components carry with them limitations.
\skye\ is not applicable in the limit of nuclear densities or temperatures: ions are treated as charged point particles and all nuclear interactions are ignored.
Several finite-temperature, composition-dependent, hot nuclear matter EOSs 
have been developed for this regime, including those based on nonrelativistic Skyrme parametrizations
\citep{lattimer_1991_aa,schneider_2017_aa}, variational approaches \citep{togashi_2017_aa} 
and relativistic mean fields \citep{sugahara_1994_aa,shen_1998_aa,typel_2010_aa,fattoyev_2010_aa,steiner_2013_aa}.

Along similar lines at low temperatures and densities, where $T \lesssim 10^{5}\,\mathrm{K}$ and $\rho \lesssim 10^3\,\mathrm{g\,cm^{-3}}$, our ion-ion interaction term becomes large and negative, resulting in unphysical results such as negative entropy.
This reflects the fact that matter is not fully ionized in this limit.
In reality bound states form, reducing the mean ion charge and so reducing the ion-ion interactions.
For very low densities this results in an ideal gas with a different mean molecular weight.
Several EOSs have been developed for this regime, including those based on 
free energy minimization \citep{Saumon1995,Irwin2004},
cluster activity expansions \citep{rogers_1974_aa,rogers_1981_aa,Rogers2002},
cluster viral expansions \citep{omarbakiyeva_2015_aa,ballenegger_2018_aa},
density-functional theory molecular dynamics \citep{militzer_2013_aa,2014ApJS..215...21B}, 
path integral Monte Carlo \citep{militzer_2001_aa}, 
quantum Monte Carlo \citep{mazzola_2018_aa},
Feynman-Kac path integral representations \citep{alastuey_2020_aa}, and
asymptotic expansions \citep{alastuey_2012_aa}.
Using these EOSs in stellar evolution calculations typically requires 
pre-tabulating results for fixed compositions due to the computational cost of solving for ionization equilibrium.

In principle partial ionization could be included in \skye\ in a variety of ways.
For instance we could add terms accounting for electron-ion interactions, but unfortunately we are not aware of robust prescriptions for the interaction free energy $F_{j-e}$ in this limit.
The challenge is that existing prescriptions are based on perturbation expansions~\citep{1961ApJ...134..669S,Potekhin2010}, but these break down well before the formation of bound states~\citep{1976PhRvA..14..816G}.
Variational approaches seem more promising in this limit, but are more computationally expensive to implement because they involve minimizing the free energy with respect to a variational parameter~\citep{1976PhRvA..14..816G}.
The same is true for direct solutions to the Saha equation, which are generally quite expensive.

A further limitation concerns our understanding of high density
quantum melts.  The physics is not as well understood as for lower
densities or higher temperatures. We think this is a fruitful area 
for further study, particularly given that the quantum melt line
\skye\ currently predicts disagrees with calculations based
on the Lindemann criterion~\citep{1993ApJ...414..695C,PhysRevB.18.3126,PhysRevLett.76.4572}.

Putting these limitations together, we recommend that \skye\ not be used for densities above $0.1 \rho_{j,\rm nuclear} \approx A_j 10^{13}\mathrm{g\,cm^{-3}}$, where $A_j$ is the number of baryons per ion, or for temperatures above the proton rest mass-energy $T_{\rm QCD} \approx 10^{13}\,\mathrm{K}$.
We further recommend that \skye\ not be used in the joint limit $T < T_{j}^{\rm ion}$ and $\rho < \rho_{j}^{\rm ion}$.
Here $T_{j}^{\rm ion}$ is the temperature above which a dilute gas is fully ionized.
Neglecting degeneracy factors, we may solve for this using the Saha equation
\begin{align}
	\frac{n_{j,Z_j}}{n_{j,Z_j-1}} = \frac{2 n_{Q,e}}{n_e} e^{-\psi_{f,j} / k_{\rm B} T},
\end{align}
where $\psi_{f,j}$ is the final ionization potential of a species of charge $Z_{j}$, and  $n_{j,Z}$ is the number density of fully ionized ions of species $j$ and charge $Z$.
As a rough heuristic we require $n_{j,Z_j} > 10 n_{n_{j,Z_j-1}}$ to ensure that full ionization is a good approximation.
With this we find
\begin{align}
	k_{\rm B} T_{j}^{\rm ion} \approx \frac{\psi_{f,j}}{\frac{3}{2} \ln (T_{j}^{\rm ion}/10^4\,\mathrm{K}) - \ln (Z_j \rho / A_j \mathrm{g\,cm^3}) - 7}.
\end{align} 
If we approximate $\psi_{f,j} \approx \mathrm{Ry} Z_j^2$ we then find
\begin{align}
	T_{j}^{\rm ion} \approx \frac{10^5\,\mathrm{K} Z_j^2}{\frac{3}{2} \ln (T_{j}^{\rm ion}/10^4\,\mathrm{K}) - \ln (Z_j \rho / A_j \mathrm{g\,cm^3}) - 7}.
\end{align} 
For densities below that of pressure ionization this typically gives $T_{j}^{\rm ion} \approx 10^4\,\mathrm{K} Z_j^2$.
Along similar lines, $\rho_{j}^{\rm ion}$ is the density above which a low-temperature system is fully ionized, given approximately by~\citep{doi:10.1098/rspa.1938.0073}
\begin{align}
	\rho_{j}^{\rm ion} &= \frac{3 m_{j}}{\pi \sqrt{2}} \left(\frac{\psi_{f,j}}{e^2 a_0}\right)^{3/2}\\
	&\approx 3 \frac{m_j}{m_p} Z_j^{3} \mathrm{g\,cm^{-3}}\approx 3 A_j Z_j^{3} \mathrm{g\,cm^{-3}},
\end{align}
where $a_0 = \hbar^2 / m_e e^2$ is the Bohr radius.
For mixtures of ions we recommend averaging $\rho_j^{\rm ion}$ and $T_j^{\rm ion}$ weighted by number density to determine the appropriate limits.
Finally, we recommend caution in interpreting results in the quantum melt limit, which occurs in the joint limit of $\rho > (A_j/12)^4 (Z_j/6)^6 10^9 \mathrm{g\,cm^{-3}}$ and $T < (A_j/12)(Z_j/6)^4 10^7\mathrm{K}$.

\section{Thermodynamics via Automatic Differentiation} \label{sec:autodiff}

\skye\ computes thermodynamic quantities from a free energy and its derivatives.
Modern stellar evolution software instruments require not only the first derivatives, which supply the energy, entropy, and pressure, but also second derivatives, which supply specific heats and susceptibilities.
Moreover because stellar evolution is often numerically stiff it is generally solved implicitly with a Newton-Raphson method.
The Jacobian of that method then requires derivatives of each of these thermodynamic quantities and so requires third derivatives of the free energy.
Because of this, the performance and convergence of stellar evolution calculations depends strongly on being able to compute high-quality derivatives of the structure equations with respect to the structure variables ($\rho, T, \{y_j\}, ...$ in each cell).
These derivatives in turn depend on derivatives from the equation of state, and so it is important that the derivatives reported by the EOS actually be derivatives of the corresponding quantities (i.e., $\partial p/\partial \rho$ should be a good approximation to the variation of $p$ with $\rho$).

To supply these derivatives we compute the analytic free energy using forward-mode operator-overloaded automatic differentiation~\citep{2000JCoAM.124..171B}.
Specifically, we define a numeric Fortran type \code{auto\_diff\_real\_2var\_order3} which contains a floating-point number as well as its first, second, and third partial derivatives with respect to two independent variables, temperature and density.
For example, if \code{x} is of this type then it contains elements \code{x\%val} representing the value of \code{x}, \code{x\%d1val1} for the value of $\partial x/\partial T|_\rho$, \code{x\%d1val2} for $\partial x/\partial \rho|_T$, \code{x\%d1val1\_d1val2} for $\partial^2 x/\partial \rho \partial T$, and so on.

This new numeric type overloads operators to implement the chain rule.
So in the code a line such as \code{f = x * y} is overloaded to set
\begin{align}
	\code{f\%val} &= \code{x\%val * y\%val} \\
	\code{f\%d1val1} &= \code{x\%d1val1 * y\%val + y\%d1val1 * x\%val} \\
	\code{f\%d1val2} &= \code{x\%d1val2 * y\%val + y\%d1val2 * x\%val} 
\end{align}
and so on.
These expressions rapidly become more complicated for higher-order derivatives, but the basic principle is the same.
We generate the overloaded operators using a \code{Python} program which computes power series using \code{SymPy}~\citep{10.7717/peerj-cs.103} and extracts chain-rule expressions.
These are then optimized to eliminate common sub-expressions and to minimize the number of division operators, and then translated into \code{Fortran}.
All of this functionality is built on top of the CR-LIBM software package \citep{CR-LIBM}, which enables bit-for-bit identical results across all platforms.

With this numeric type, modifying the \skye\ free energy is simple: translate analytic formulas into \code{Fortran}.
Additional terms such as
\begin{align}
	\delta F = k \rho e^{T / \sqrt{\rho}} \rightarrow \code{k * rho * exp(T / sqrt(rho))}
\end{align}
can be written as-is, and all derivatives are provided automatically.

We have developed further machinery to support derivatives with respect to a variable number of ion abundances, built using the parameterized derived type feature of  \code{Fortran} 2003.
Unfortunately compiler support for this feature is lacking, and neither \code{gfortran} v10.2.0 nor \code{ifort} v19.0.1.144 fully implement it.
Future \code{Fortran} compilers may implement this feature, at which point \skye\ will be able to provide derivatives with respect to composition in addition to the usual $\rho$ and $T$ derivatives.

\section{Applications}\label{sec:examples}

We now explore the properties of \skye\ and compare it with PC EOS and HELM EOS.
When we refer to PC and HELM in the following we mean the MESA implementation
of each.  For PC this is based on source code made available by
A.~Potekhin.  It has been modified during its incorporation into MESA,
but not in ways that intentionally affect its results except for a
numerical blurring of the Coulomb phase transition. Likewise, the original source code of HELM
has been modified during its incorporation into MESA. Examples of such modifications include
providing third derivatives of the Helmholtz free energy and second derivatives
of the electron chemical potential,
using more accurate quadrature summations for derivatives of the Fermi-Dirac functions
when forming derivatives of the Helmholtz free energy \citep{gong_2001_aa},
supplying denser tables of the Helmholtz free energy and eight of its partial derivatives
(100 point per decade grid densities in $\rho$ and $T$), 
adding controls to activate or deactivate the pieces of physics in HELM, 
and deploying CR-LIBM \citep{CR-LIBM} for an efficient and proven correctly-rounded mathematical library
to ensure bit-for-bit identical results across platforms.

\subsection{Derivative Quality}

\begin{figure}
	\centering
	\includegraphics[width=0.45\textwidth]{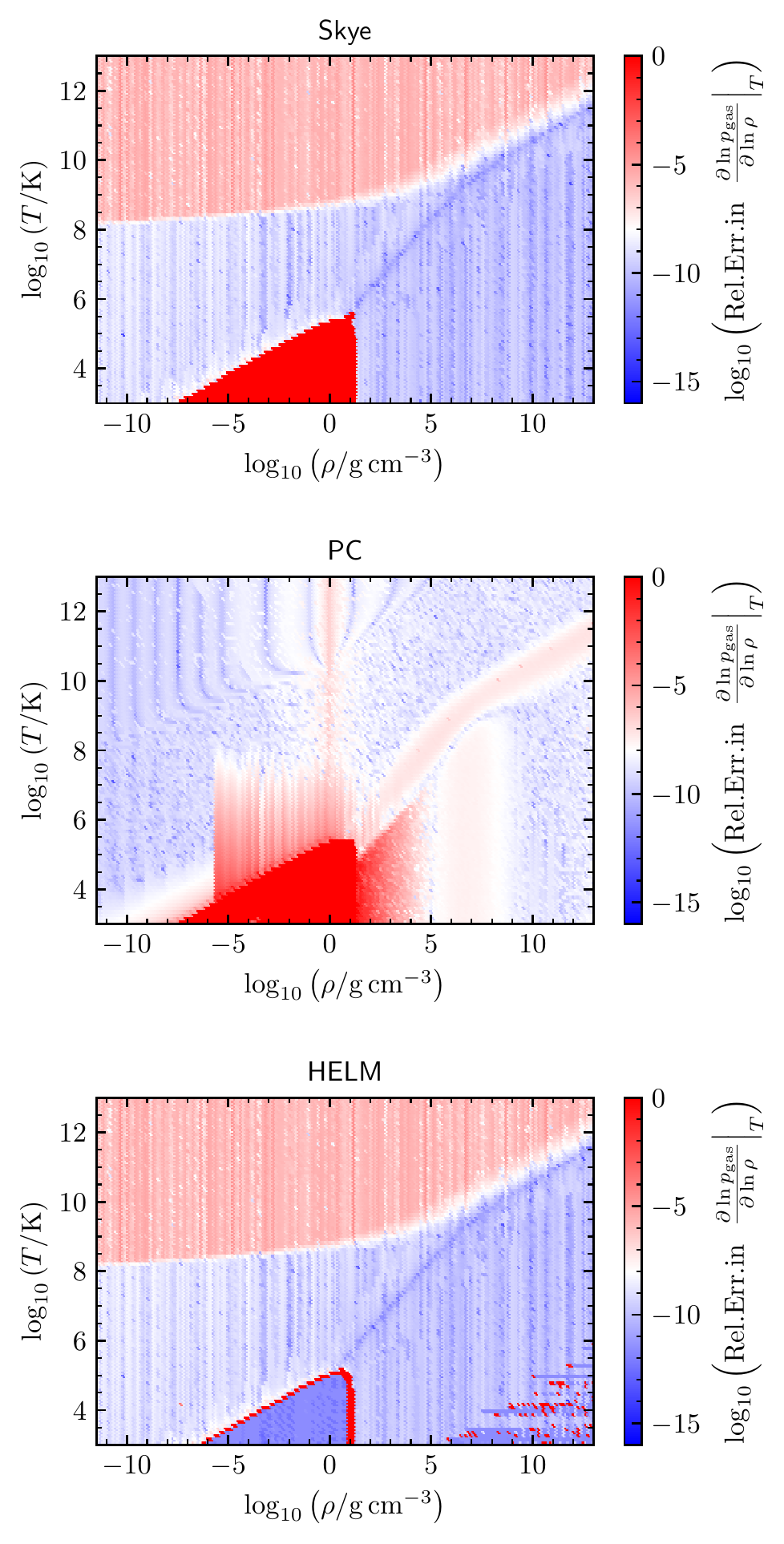}
	\caption{The logarithm of the relative difference between $\partial \ln p_{\rm gas}/\partial \ln \rho|_T$ and a finite difference approximation to the same is shown as a function of $T$ and $\rho$ for each of \skye, PC, and HELM for an equal-mass fraction mixture of $^{12}\mathrm{C}$ and $^{16}\mathrm{O}$. The feature in \skye\ and PC at intermediate densities and low temperatures results from negative pressures caused by the assumption of a fully ionized free energy in a region that should form bound states, indicating that these EOSes are not valid in that limit.}
	\label{fig:dlnPgas_dlnRho}
\end{figure}

Figure~\ref{fig:dlnPgas_dlnRho} shows the relative difference between the reported derivative $\partial \ln p_{\rm gas}/\partial \ln \rho|_T$ and an iteratively acquired high-precision numerical derivative \citep[e.g.,][]{ridders_1982_aa,press_1992_aa} for each of \skye, HELM and PC.
Here $p_{\rm gas}$ is the total pressure minus radiation pressure.
For HELM and \skye\ we used the directly reported partial derivative while for PC we used $\partial \ln p_{\rm gas}/\partial \ln \rho|_T = \chi_\rho$.

Both \skye\ and HELM produce high-quality derivatives, better than one part in $10^{8}$, over much of the $\rho-T$ plane.
This is because \skye\ uses automatic differentiation on the analytic portion of the free energy and both \skye\ and HELM use spline partial derivatives on the tabulated ideal electron-positron free energy, so the quality of derivatives of thermodynamic quantities in these equations of state is limited only by the precision of floating-point arithmetic.
The PC derivative quality is somewhat lower than this primarily because of an internal redefinition of the density which occurs in the code but which is not propagated through the subsequent derivatives.

The grid structure in the derivative quality is set by the spacing of the HELM ideal electron-positron free energy table, on which both \skye\ and HELM rely.
At high temperatures above $10^9\,\mathrm{K}$ the system becomes dominated by electron-positron pairs and so nearly independent of the $\rho$.
The derivatives are then pushed towards the limits of floating point precision, degrading their quality.

The feature in \skye\ and PC at intermediate densities ($\rho \sim 1\,\mathrm{g\,cm^{-3}}$) and low temperatures ($T < 10^5\,\mathrm{K}$) results from negative pressures caused by the assumption of a fully ionized free energy in a region that should form bound states,
indicating that these equations of state are not valid in that limit.

\begin{figure}
	\centering
	\includegraphics[width=0.45\textwidth]{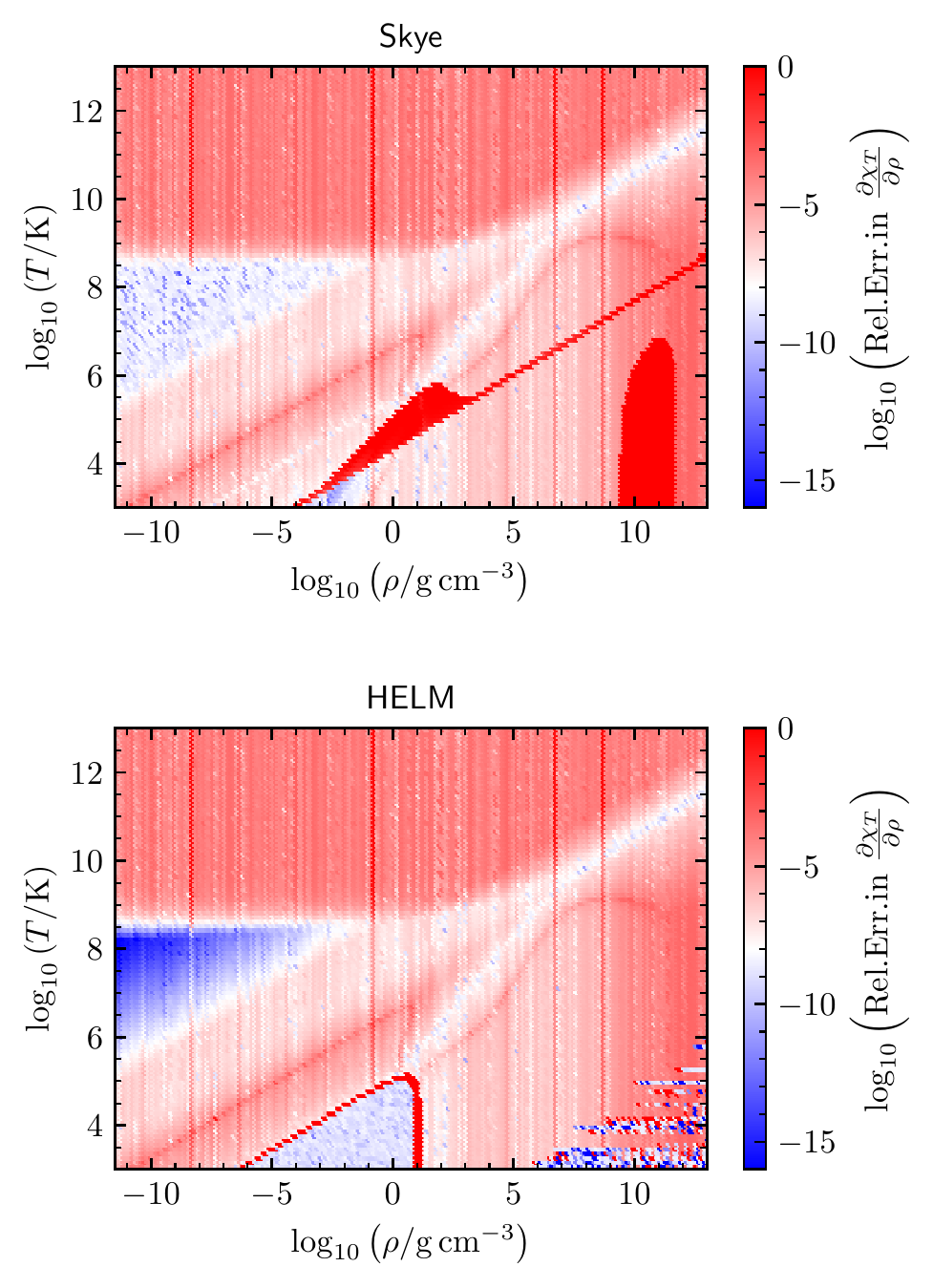}
	\caption{The logarithm of the relative difference between $\partial \chi_T/\partial \ln \rho|_T$ a finite difference approximation to the same is shown as a function of $T$ and $\rho$ for \skye\ and HELM for an equal-mass fraction mixture of $^{12}\mathrm{C}$ and $^{16}\mathrm{O}$.}
	\label{fig:dChi_dlnRho}
\end{figure}

In general the quality of derivatives degrades as we look to higher orders because there is more room for precision issues.
Figure~\ref{fig:dChi_dlnRho} shows the relative difference between the reported derivative $\partial \chi_T/\partial \ln \rho|_T = \partial^2 \ln p / \partial \ln \rho\,\partial \ln T$ and an iteratively acquired high-precision numerical derivative for \skye\ and HELM.
Once more at high temperatures above $10^9\,\mathrm{K}$ the system becomes dominated by electron-positron pairs and so nearly independent of the $\rho$.
The derivatives in \skye\ and HELM are then pushed towards the limits of floating point precision, degrading their quality.

$\partial \chi_T/\partial \ln \rho|_T$ is not reported natively by PC so we could not include PC in this comparison.
Because MESA requires this derivative, when PC is used in MESA this derivative is estimated using finite differences in $\ln \rho$.  This results in derivatives that are accurate at only around the $10^{-2}$ level, which was often a bottleneck in stellar evolution calculations.

\subsection{Thermodynamic Consistency}

The first law of thermodynamics is an exact differential and thus implies several consistency relations between the different thermodynamic quantities.
These are~\citep[][see their Appendix A.1.3]{Timmes2000,Paxton2019}
\begin{align}
	\mathrm{dpe} \equiv \frac{\rho^2}{p}\left.\frac{\partial e}{\partial \rho}\right|_{T,\{y_j\}} + \frac{T}{p}\left.\frac{\partial p}{\partial T}\right|_{\rho,\{y_j\}} - 1 &= 0\\
	\mathrm{dse} \equiv T\frac{\partial s/\partial T|_{\rho,\{y_j\}}}{\partial e/\partial T|_{\rho,\{y_j\}}} - 1 &= 0\\
	\mathrm{dsp} \equiv -\rho^2\frac{\partial s/\partial \rho|_{T,\{y_j\}}}{\partial p/\partial T|_{\rho,\{y_j\}}} - 1 &= 0.
\end{align}
If these relations are not satisfied an equation of state is thermodynamically inconsistent.
For simulations of physical scenarios this can result in artificial generation or loss of energy or entropy or incorrect conversion between these and mechanical work.
Moreover thermodynamic inconsistency means that different forms of the same physical equations are not even mathematically identical.
For instance, neglecting changes in composition, in stellar evolution the equation of local energy conservation is often written as~\citep{Paxton2015}
\begin{align}
	\frac{de}{dt} - \frac{p}{\rho}\frac{d\ln \rho}{dt} = T \frac{ds}{dt},
\end{align}
or alternatively as
\begin{align}
	c_p T \left[(1-\nabla_{\rm ad} \chi_T) \frac{d\ln T}{dt} - \nabla_{\rm ad} \chi_\rho \frac{d\ln \rho}{dt}\right] = T \frac{ds}{dt}.
\end{align}
For numerical reasons it is often preferable to use one form over another, but these forms are only mathematically equivalent to the extent that the EOS is thermodynamically consistent.

\begin{figure}
	\centering
	\includegraphics[width=0.45\textwidth]{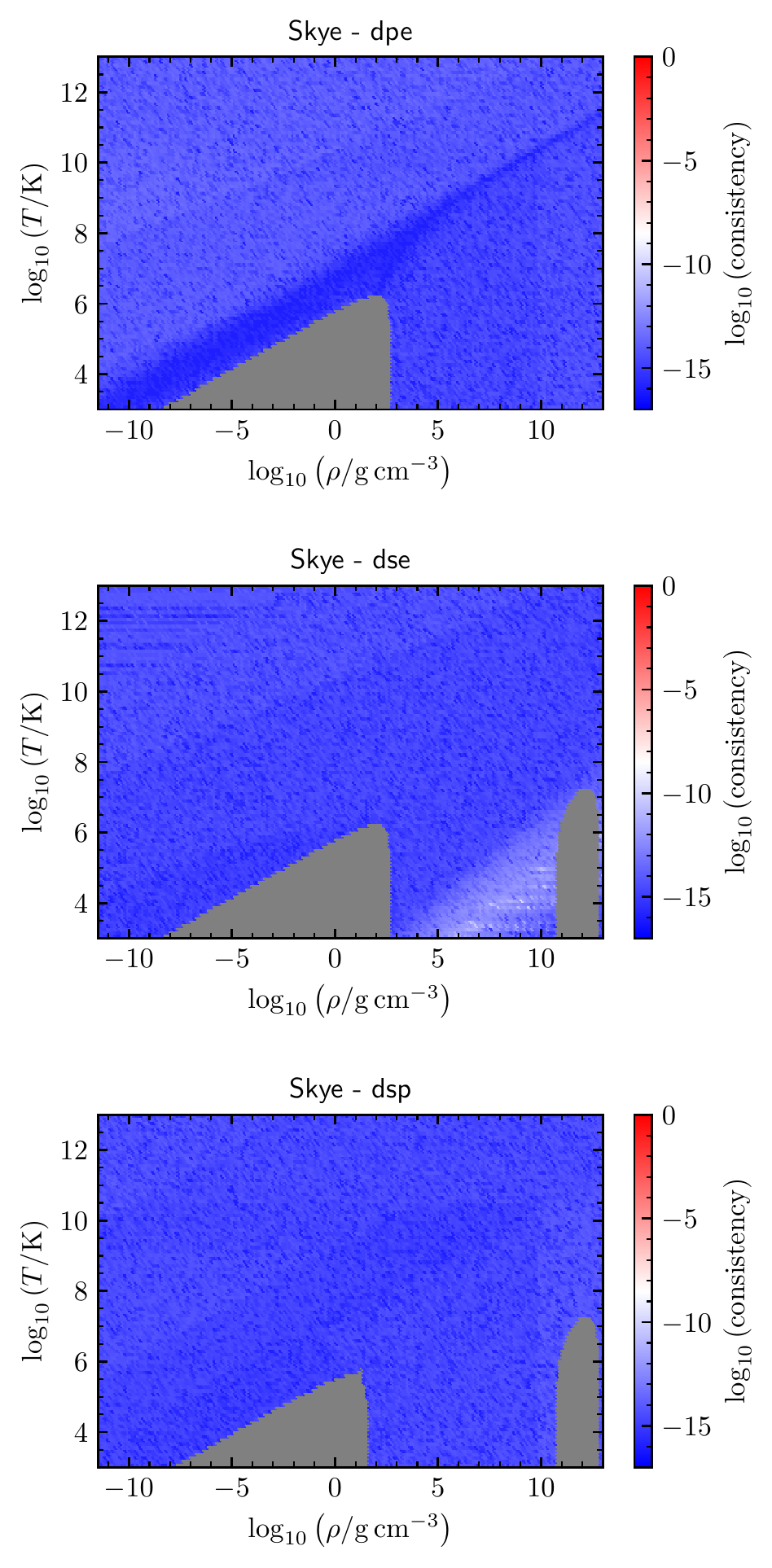}
	\caption{The thermodynamic consistency measures $\mathrm{dpe}$, $\mathrm{dse}$, and  $\mathrm{dsp}$ are shown for \skye\ on a logarithmic scale for an equal-mass fraction mixture of $^{16}\mathrm{O}$ and $^{20}\mathrm{Ne}$. Grey indicates regions where the result is NaN due to negative reported entropy, energy, or pressure (solid regions) resulting in undefined logarithms in intermediate steps of the calculation. The feature at intermediate densities and low temperatures indicates negative pressures caused by our assumption of a fully ionized free energy in a region that should form bound states, indicating that \skye\ is not valid in that limit.}
	\label{fig:consistency}
\end{figure}

Figure~\ref{fig:consistency} shows the quantities $\mathrm{dpe}$,  $\mathrm{dse}$, and  $\mathrm{dsp}$ from \skye\ as functions of $\rho$ and $T$ for an equal-mass fraction mixture of $^{16}\mathrm{O}$ and $^{20}\mathrm{Ne}$.
Because \skye\ is derived from a free energy formalism it is thermodynamically consistent to the limits of floating-point precision.

Note that this high degree of consistency should not be confused with physical accuracy.
\skye\ returns \emph{numerically accurate} partial derivatives and thermodynamically consistent quantities, but this is not the same as \emph{physical} accuracy, which is a matter of how well the input physics matches Nature.

\subsection{Crystallization Curves}
\label{sec:ccurve}

\begin{figure}
	\centering
	\includegraphics[width=0.45\textwidth]{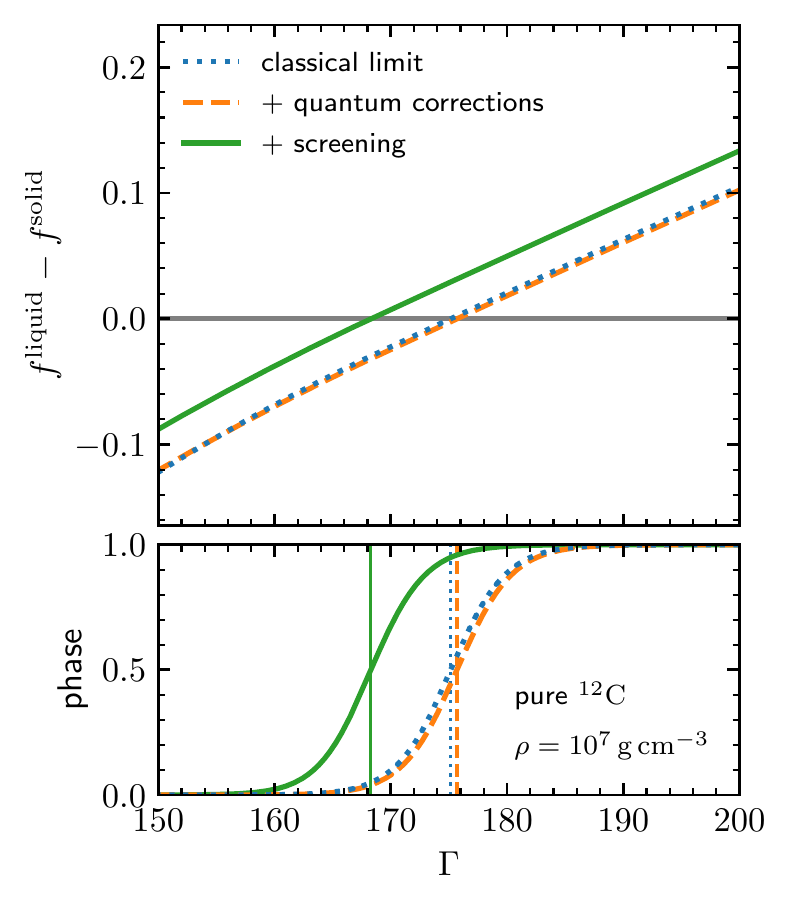}
	\caption{The liquid-solid free energy difference (top panel) and phase $\phi$ (bottom panel) as a function of $\Gamma$ for pure $^{12}\rm C$ plasma at $\rho=10^7\mathrm{g\,cm^{-3}}$.  We show the effects of different terms in the free energy by first showing the result in the classical limit (forcing $\eta \to 0$), then adding quantum effects, and finally including screening corrections.}
	\label{fig:phase_OCP}
\end{figure}

We demonstrate where and how crystallization occurs in \skye\ by first
considering a pure $^{12}\rm C$ plasma at
$\rho=10^7\mathrm{g\,cm^{-3}}$.  Figure~\ref{fig:phase_OCP} shows the
location of crystallization and how that depends on which terms are
included in the free energy.\footnote{We can ignore any terms in the
  free energy that are not phase-dependent.}  The dotted line shows
the result of considering only the classical OCP free energy, which we
achieve by artificially forcing $\eta \to 0$ and deactivating the
screening terms.  This illustrates that crystallization is centered at
the established value of $\Gamma \approx 175$
\citep[e.g.,][and references therein]{PhysRevE.62.8554} and occurs over an an interval of
width $\Delta \Gamma \approx 10$ due to the blur described in
Section~\ref{sec:thermo}.  Including quantum corrections causes a
small shift ($\delta \Gamma \lesssim 1$) to higher values of $\Gamma$.
Adding screening results in much larger shift
($\delta \Gamma \approx 7$) towards lower values of $\Gamma$.%
\footnote{The size of this shift is larger than is shown in
  Figure 7 of \citet{PhysRevE.62.8554}. That calculation was done without quantum effects
  and used the fit from \citet{Yakovlev1989} for screening in the liquid regime instead of using Equation~(19) in \citet{PhysRevE.62.8554}.  The values of $f_{\rm ie}$ according to the two expressions are very close: the difference is $< 2$\%.  However, this difference in the screening correction is sufficient to noticeably affect the $\Gamma$ at which crystallization occurs, highlighting the sensitivity of the liquid/solid phase transition in Coulomb plasmas to tiny details in the free energy.}

\begin{figure}
	\centering
	\includegraphics[width=0.45\textwidth]{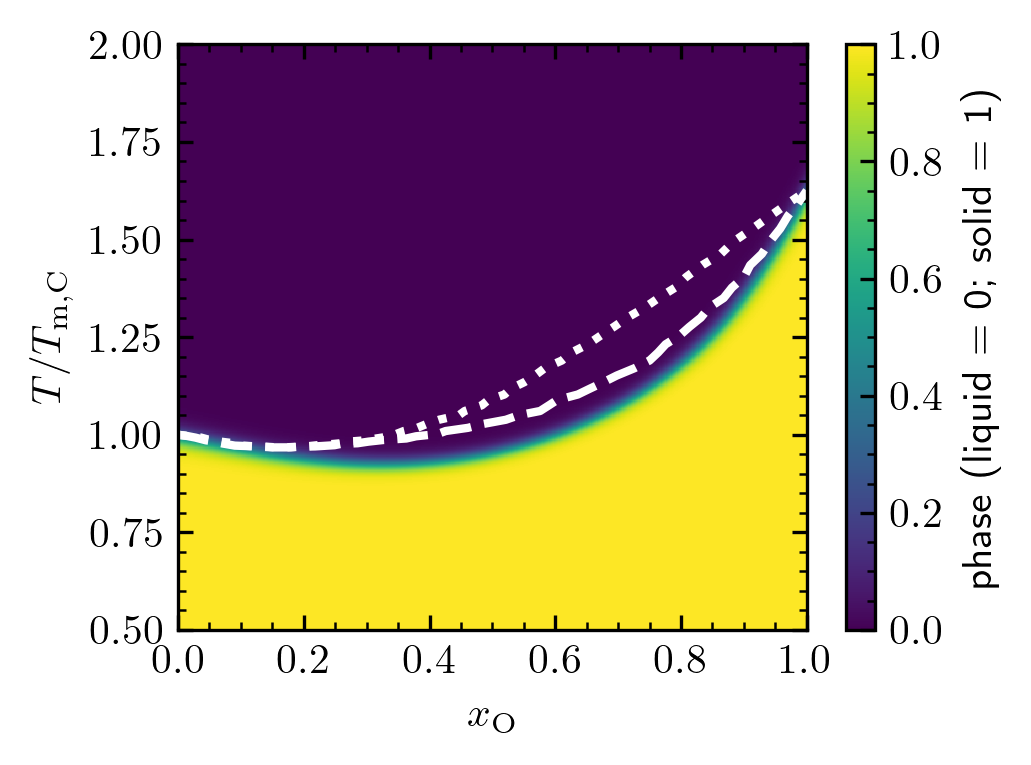}
	\caption{The phase $\phi$ as a function of the ratio of the temperature to the melting temperature of a pure $^{12}\rm C$ plasma, $T/T_{\rm m,C}$, and $^{16}\rm O$ number fraction, $x_{\rm O}$, at fixed density $\rho=10^7\mathrm{g\,cm^{-3}}$ for a mixture of $^{12}\rm C$ and $^{16}\rm O$.  The white lines are the liquidus (dotted) and solidus (dashed) from \citet{Blouin2020}.}
	\label{fig:phase_PC13}
\end{figure}

\skye\ determines the phase (solid/crystalline or liquid) self-consistently via free energy minimization, so it can model the effects of varying composition on melting temperature.
Figure~\ref{fig:phase_PC13} shows the phase as a function of temperature and composition in a $^{12}{\rm C}$-$^{16}{\rm O}$ mixture.
The x-axis, $x_{\rm O}$, is the $^{16}\rm O$ number fraction.  The y-axis, $T/T_{\rm m,C}$, is the ratio of the temperature to the melting temperature of a pure $^{12}\rm C$ plasma.
Because $\phi$ is a smoothed measure of the phase it takes a non-zero width to transition from $\phi \approx 0$ to $\phi \approx 1$.

The work of \citet{Blouin2020}, which adopts a Gibbs–Duhem integration technique coupled to Monte Carlo simulations, provides a useful point of comparison.
Their phase curve is calculated at $P = 10^{24}\,\mathrm{erg\,cm^{-3}}$ and so we calculate the Skye phase at $\rho=10^7\,\mathrm{g\,cm^{-3}}$ which corresponds to a similar pressure of $P \approx 8 \times 10^{23}\,\mathrm{erg\,cm^{-3}}$.
In Figure~\ref{fig:phase_PC13}, we show the \citet{Blouin2020} liquidus and solidus.
The reference melting temperature used for the \citet{Blouin2020} liquidus and solidus curves is the $T_{\rm m,C}$ value from \citet{Blouin2020}, which differs from the Skye value.
Recall Skye does not  consider phase separation, so it produces a single (blurred) transition line.

\begin{figure}
	\centering
	\includegraphics[width=0.45\textwidth]{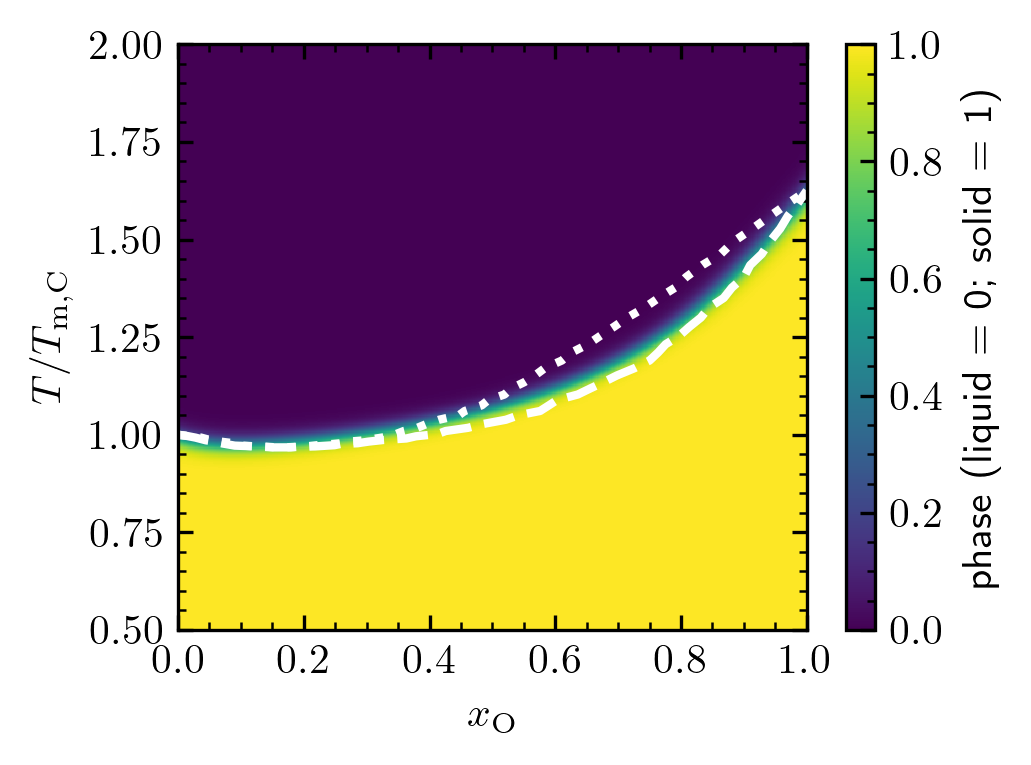}
	\caption{Same as Figure~\ref{fig:phase_PC13}, except replacing the default solid mixing free energy from \citet{2013A&A...550A..43P} with the form proposed by \citet{1993PhRvE..48.1344O}.}
	\label{fig:phase_O93}
\end{figure}

As an example of how simple it is to swap out individual components in
the \skye\ framework, Figure~\ref{fig:phase_O93} shows the result
when we replace the (default) fit of \citet{2013A&A...550A..43P} for the solid mixing corrections with the form proposed by \citet{1993PhRvE..48.1344O}.
The \citet{2013A&A...550A..43P} form is in part motivated to overcome unphysical behavior%
\footnote{Specifically, \citet{2013A&A...550A..43P} note that the
  Ogata function is non-monotonic for fixed $x_2 < 0.5$ at $R > 2$.
  This is not simply a misbehaving fit.  The values in Table II of
  \citet{1993PhRvE..48.1344O} that are being fit show the same
  non-monotonic behavior.}
present in the \citet{1993PhRvE..48.1344O} fit at charge ratios $R > 2$,
though a C/O mixture ($R = 4/3$) is not in the troublesome regime.

The agreement shown in Figure~\ref{fig:phase_O93} between
\citet{Blouin2020} and Skye when using
\citet{1993PhRvE..48.1344O} is anticipated.
The results of \citet{Blouin2020} agree well with the results of
\citet{2010PhRvE..81c6107M}.  In turn, \skye\ resembles the
analytic-fit-based approach of \citet{2010PhRvE..81c6107M}, with the same extension from two-component to multi-component plasmas, and \citet{2010PhRvE..81c6107M} uses the \citet{1993PhRvE..48.1344O} formulation of the solid mixing free energy.

\begin{figure}
  \centering
  \includegraphics[width=0.45\textwidth]{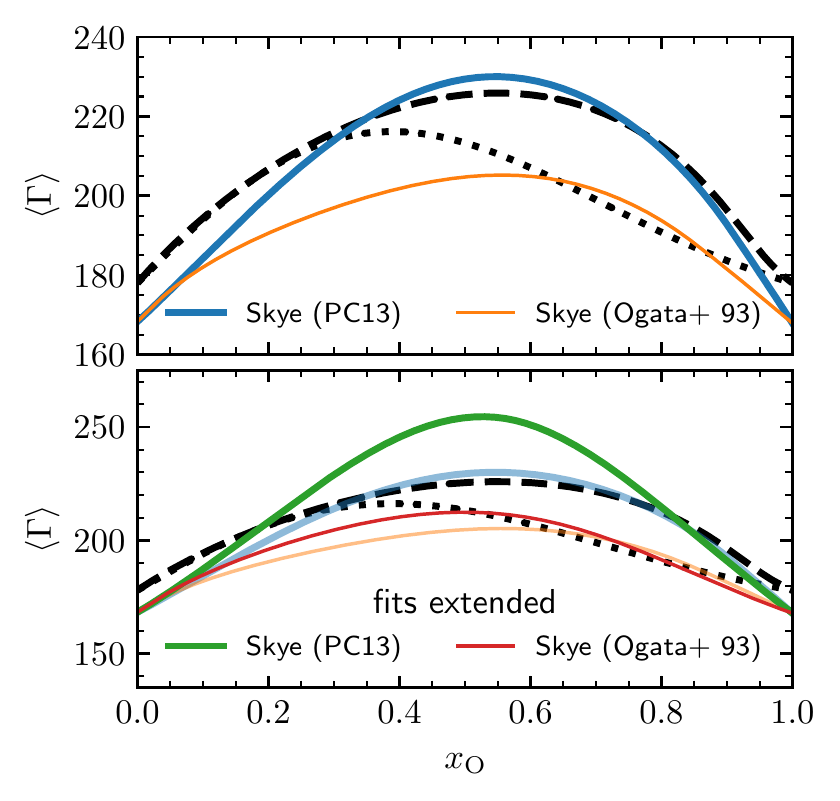}
  \caption{Value of $\langle \Gamma \rangle$ corresponding to the center of the Skye phase transition
    ($\phi = 0.5$) as function of
    $^{16}\rm O$ number fraction $x_{\rm O}$ at fixed density
    $\rho=10^7\,\mathrm{g\,cm^{-3}}$ for a mixture of $^{12}\rm C$ and
    $^{16}\rm O$.
    The top panel compares the Skye results using the indicated solid mixing correction.
    The bottom panel shows the results when extending the limits of the solid and liquid fits (see text).
    The curves from the top panel are faintly shown for ease of comparison.
    In both panels, the black lines are the liquidus (dotted) and solidus (dashed) from \citet{Blouin2020}.}.\label{fig:phase_gamma}
\end{figure}

The results shown in Figures~\ref{fig:phase_PC13} and \ref{fig:phase_O93}
are summarized in the top panel of Figure~\ref{fig:phase_gamma} which plots the value
of the average Coulomb parameter $\langle \Gamma \rangle$ at
crystallization (defined as when $\phi = 0.5$) as a function of the
$^{16}\rm O$ number fraction.  For pure compositions, the Skye phase transition
occurs at a $\langle \Gamma \rangle$ value of about 10 less than \citet{Blouin2020},
primarily reflecting the screening corrections shown in Figure~\ref{fig:phase_OCP}.
The two approaches to the mixing corrections give significantly
different values for the phase transition in an equal (by number) mixture, with the \citet{1993PhRvE..48.1344O} form yielding
$\langle \Gamma \rangle \approx 205$ and the \citet{2013A&A...550A..43P} form yielding
$\langle \Gamma \rangle \approx 230$, with the \citet{Blouin2020} results intermediate.

%
% We see that, with the
% \citet{1993PhRvE..48.1344O} solid mixing corrections, crystallization in \skye\
% for a C/O binary mixture occurs within a $\langle \Gamma \rangle$ difference
% $\lesssim 20$ from the \citet{Blouin2020} liquidus for all
% compositions.
%
% The disagreement with the results of \citet{Blouin2020} caused by
% using the \citet{2013A&A...550A..43P} fit shown in
% Figures~\ref{fig:phase_PC13} and \ref{fig:phase_gamma} leads us to
% select the \citet{1993PhRvE..48.1344O} fit, especially since early
% applications of \skye\ will be in white dwarf cooling
% (Section~\ref{sec:cooling}).
% %
% At present, work focused on high-charge ratio mixtures ($R > 2$) might make a
% different selection.
% In particular the formula of~\citet{2013A&A...550A..43P} was constructed to match that of~\citet{2003CoPP...43..279D} at low charge ratios while exhibiting more physical behavior at high charge ratios.
%
Because the range of $\Gamma_j$ where both the liquid and solid free
energy fits are valid is small, for charge ratios greater than
$(\Gamma_{\rm max}^{\rm liquid}/\Gamma_{\rm min}^{\rm solid})^{1/2}
\approx 1.1$ one species or the other will typically be extrapolated
at the phase transition.  To illustrate this effect, the bottom panel
of Figure~\ref{fig:phase_gamma} shows a `fits extended' calculation
where we used $\Gamma_{\rm min}^{\rm solid}=100$ and
$\Gamma_{\rm max}^{\rm liquid} = 300$.  This shows that the
transition $\langle\Gamma\rangle$ may depend at the 10~per-cent level
on the choice of $\Gamma_{\rm max}^{\rm liquid}$ and $\Gamma_{\rm min}^{\rm solid}$.

The structure of \skye\ demands individual fits
that behave well over wide parameter ranges and a set of prescriptions
that can collectively work well together.
This is especially necessary for determining the location of the phase
transition, given the small relative difference between the liquid and
solid free energies.
We observe that Skye, at some unusual conditions, 
reports that material returns to the liquid state at sufficiently low
temperature as a result of the quantum corrections.
We discuss this behavior in Appendix~\ref{sec:quantum}.
We hope that the ease of
experimentation with \skye\ can help motivate improved fits for some
of the key quantities.

\subsection{Comparison with Other EOS}

\begin{figure}
	\centering
	\includegraphics[width=0.45\textwidth]{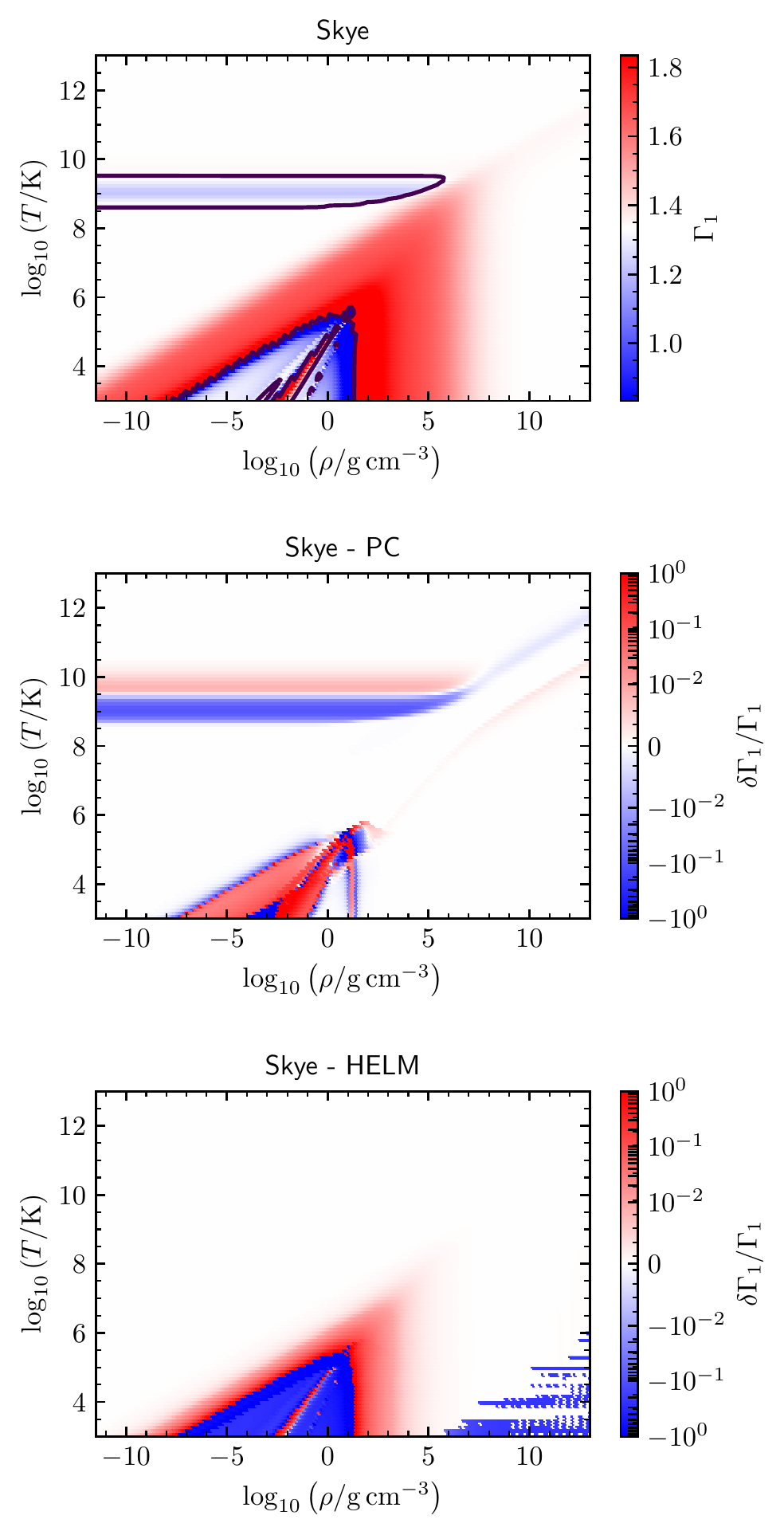}
	\caption{The adiabatic index $\Gamma_1$ is shown as a function of temperature and density for each of \skye, PC, and HELM for an equal-mass fraction mixture of $^{12}\mathrm{C}$ and $^{16}\mathrm{O}$. The upper panel is for \skye\, the others show the relative difference between \skye\ and PC and HELM. The outlined contour shows where $\Gamma_1 = 4/3$, signalling onset of the pair production instability. Note that the outlined region in the bottom-right has $\Gamma_1 < 4/3$ because of precision issues in the ideal electron-positron tables and is not a sign of a physical instability.}
	\label{fig:gam1}
\end{figure}

We now compare various outputs from \skye, PC and HELM.
Figure~\ref{fig:gam1} shows the adiabatic index $\Gamma_1$ as a function of $\rho$ and $T$ for an equal-mass fraction mixture of $^{12}\mathrm{C}$ and $^{16}\mathrm{O}$.
The upper panel is for \skye\, the others show the signed logarithm of the relative difference between \skye\ and PC and HELM.
The outlined contour shows where $\Gamma_1 = 4/3$, signalling onset of the pair production instability.

At high temperatures and low densities ($T / 10^4 \,\mathrm{K} > (\rho/10^{-10}\,\mathrm{g\,cm^{-3}})^{1/3}$), \skye\ and HELM agree to better than one part in $10^5$, and both differ from PC by including positrons, which produce the feature that runs across the figure near $10^9\,\mathrm{K}$.

At lower temperatures and higher densities \skye\ and PC generally agree to better than one part in $10^{3}$.
The first exception is at intermediate densities and low temperatures, where both \skye\ and PC show artifacts caused by the assumption of a fully ionized free energy in a region that should form bound states, indicating these equations of state are not valid in that limit.
The other major difference is a series of scars at extreme densities and very low temperatures, which \skye\ inherits from the ideal electron-positron term in HELM.
In that regime computing thermodynamic quantities often requires subtracting very similar numbers, resulting in loss of precision.
The analytic fits PC uses for the ideal electron gas avoid this issue and produce smooth results there.

\begin{figure}
	\centering
	\includegraphics[width=0.45\textwidth]{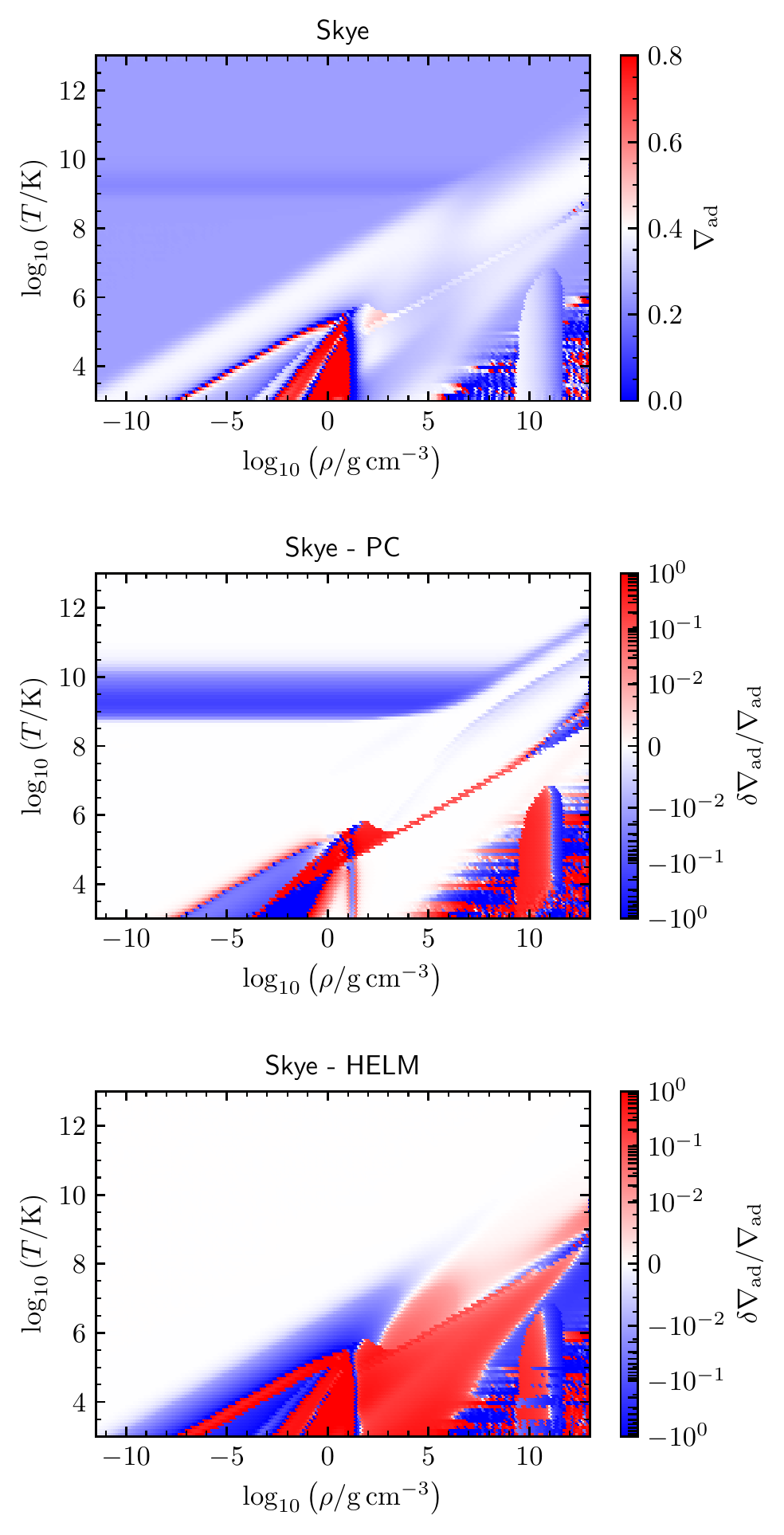}
	\caption{The adiabatic temperature gradient $\nabla_{\rm ad}$ is shown as a function of temperature and density for each of \skye, PC, and HELM for an equal-mass fraction mixture of $^{12}\mathrm{C}$ and $^{16}\mathrm{O}$. The upper panel is for \skye\, the others show the relative difference between \skye\ and PC and HELM.}
	\label{fig:grada}
\end{figure}

Closely related to $\Gamma_1$, and of particular interest for asteroseismology, is the adiabatic temperature gradient $\nabla_{\rm ad}$.
Figure~\ref{fig:grada} shows $\nabla_{\rm ad}$ as a function of $\rho$ and $T$ for the same composition used in Figure~\ref{fig:gam1}.
Once more at high temperatures \skye\ and HELM agree at the $10^{-5}$ level, and both differ from PC by including positrons.
At lower temperatures we see an order-unity difference between \skye\ and PC which stretches along a line of nearly constant $\langle \Gamma \rangle$.
This difference is because PC places the phase transition at a fixed location in $\langle \Gamma \rangle$ while \skye\ determines the phase boundary from the input physics, which in this instance causes it to place the boundary at a slightly different $\langle \Gamma \rangle$.
The other major difference is that \skye\ again shows scars at very high density that come from loss of precision in the ideal electron-positron term in HELM.
Other than that region \skye\ and PC generally agree to better than one percent.

\begin{figure}
	\centering
	\includegraphics[width=0.45\textwidth]{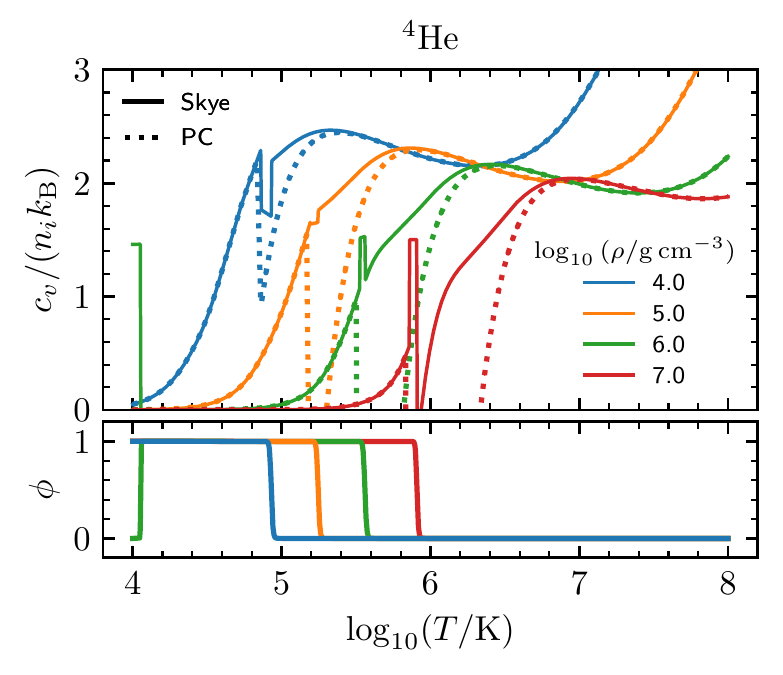}
        \includegraphics[width=0.45\textwidth]{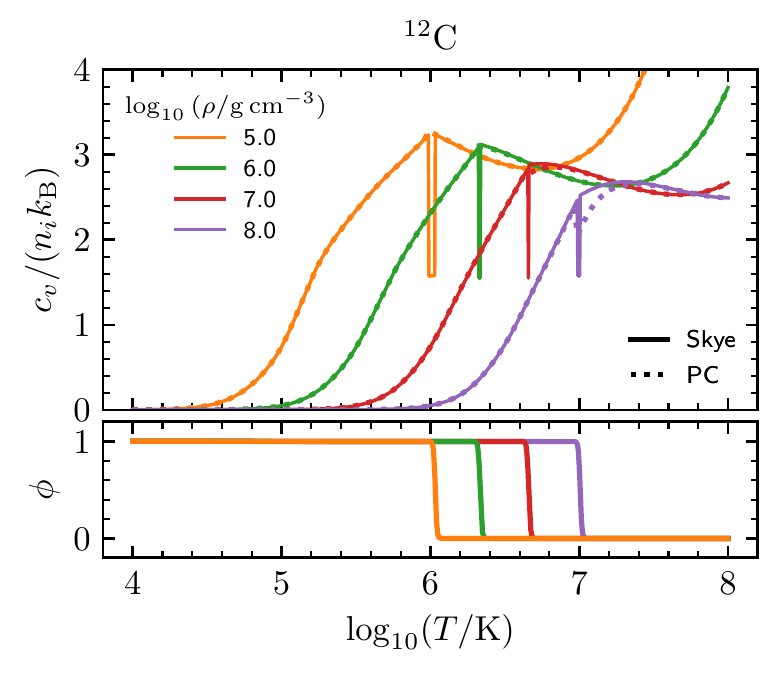}
	\caption{Comparison of specific heat between Skye and PC for $^4\rm He$ (top plot) and $^{12}\rm C$ (bottom plot) as function of temperature at an indicated set of densities.  Within each plot, the top panel shows the total specific heat at constant volume per ion for each of Skye and PC, while the bottom panel shows the Skye phase.}
	\label{fig:cv}
\end{figure}

One of the most important quantities for white dwarf cooling models
(Section~\ref{sec:cooling}) is the specific heat.  Figure~\ref{fig:cv}
compares this quantity between PC and Skye for $^4\rm He$ and
$^{12}\rm C$.  In the case of $^{12}\rm C$, the agreement is generally
good, with some disagreement at the $\approx 10\%$ level around the
temperature of crystallization in the highest density line shown.
The dips near the location of the \skye\ phase transition are due to thermodynamic extrapolation and will be discussed in detail in Section~\ref{sec:cooling}.

% This plot is all of Cv (not just the ions), which I think is why one
% doesn't see the nice 3/2 limit in the highest T/lowest Rho.

In the case of $^{4}\rm He$, in coolest parts of the liquid regime, PC
produces specific heats that fall rapidly with decreasing temperature
and even become negative.  This reflects a difference in the assumed
physics.  This version of PC contains only the leading-order term in
the Wigner-Kirkwood expansion (which is pushed beyond its range
of validity of $\eta_j \la 1$ in these plots), while \skye\ includes the prescription of
\citet{2019MNRAS.490.5839B} which is valid up to $\eta_j \approx 12$.
At densities $\log_{10}(\rho/{\rm g\,cm^{-3}}) = $ 4, 5, and 6,
Figure~\ref{fig:cv} illustrates that the \citet{2019MNRAS.490.5839B}
prescription reasonably joins onto the $\propto T^3$ specific heat of the
Debye regime. For higher $^{4}\rm He$ densities, this join becomes
less smooth and by $\log_{10}(\rho/{\rm g\,cm^{-3}}) = 7$, \skye\ too
develops regions of negative specific heat, because by $\rho \ga 10^{7}\mathrm{g\,cm^{-3}}$ for Helium, $r_{s,j} \la 300$ which is beyond the validity of the fit by~\citet{2019MNRAS.490.5839B}.
Eliminating these features awaits future improvements in prescriptions for the free energy of the
quantum Coulomb liquid.

\subsection{White Dwarf Cooling Curves} \label{sec:cooling}

We have computed white dwarf (WD) cooling curves using the Modules for
Experiments in Stellar Astrophysics~\citep[MESA;][]{Paxton2011,
  Paxton2013, Paxton2015, Paxton2018, Paxton2019} software
instrument. MESA uses a blend of several equations of state, and we
have configured the blend to use \skye\ in regions of high density or
temperature. 
Details of MESA, the blend, and other microphysics inputs are provided in Appendix~\ref{appen:mesa}.

Our example WD model is 0.6~$M_\odot$ with a C/O core and an initial
hydrogen layer mass of $5 \times 10^{-5}\ M_\odot$. This model is
based on the MESA test case \verb|wd_cool_0.6M| from MESA release
version 15140. Our cooling tracks begin when the model has a core
temperature of $\log_{10}(T_{\rm c}/K) = 7.8$ and luminosity of
1~$L_\odot$, and the WD cools until the core temperature reaches
$\log_{10}(T_{\rm c}/K) = 6.0$. We use the DA WD atmosphere tables of
\cite{Rohrmann12} as our outer boundary conditions for these WD
cooling models. The prior evolution of the WD progenitor model
included heavy element sedimentation so that the envelope is
stratified and the outer layers are composed of pure hydrogen, but for
simplicity we turn diffusion off for the cooling tracks calculated in
this paper. These models therefore do not include any cooling delay
associated with heating from sedimentation of $^{22}$Ne such as that
described in \cite{Paxton2018} and \cite{Bauer2020}. Instead, we focus
on cooling effects directly associated with EOS quantities such as
heat capacity and latent heat released by crystallization.

We run several versions of the WD cooling model described above, using
either \skye\ or the PC EOS in the high density regime
($\log_{10}(\rho/{\rm g\,cm^{-3}}) > 4$).
The PC EOS provides thermodynamics
for both liquid and solid states, with the location of the phase
transition a free parameter to be set by the user. As a baseline model
for comparison, we run the cooling WD with crystallization in PC set to
occur when the plasma reaches $\langle \Gamma \rangle = 230$, but with
no latent heat included in the model.
Previous WD cooling models using MESA have adopted this choice
of $\langle \Gamma \rangle = 230$ as a rough approximation of the C/O
phase curve in mixtures relevant for WD interiors (see
\cite{Bauer2020} for a recent example and further discussion).
We also run the WD cooling model using PC with crystallization
occurring at $\langle \Gamma \rangle = 175$ and $\langle \Gamma \rangle = 230$ with
the latent heat included in the models. 
When running with the PC EOS,
MESA models include the latent heat by taking the difference of entropy $s$
in the solid and liquid states, smoothed over a narrow range of
$\Gamma$ around the phase transition (e.g.\ in our case $228 < \Gamma < 232$
for crystallization at $\langle \Gamma \rangle = 230$). The latent
heating term is then constructed as
$\epsilon_{\rm latent} = -T (s_{\rm solid} - s_{\rm liquid})/\delta t$,
where $\delta t$ is the timestep. This latent heat is
included in the evolution as part of
$\epsilon_{\rm grav} \equiv -T ds/dt$ \citep{Paxton2018}. Finally, we
run the same WD cooling model with \skye\ as the EOS,
which includes the phase transition and the latent heat according
the phase curves shown in Figures~\ref{fig:phase_PC13}
and~\ref{fig:phase_gamma}.

\begin{figure}
  \centering
  \includegraphics[width=0.45\textwidth]{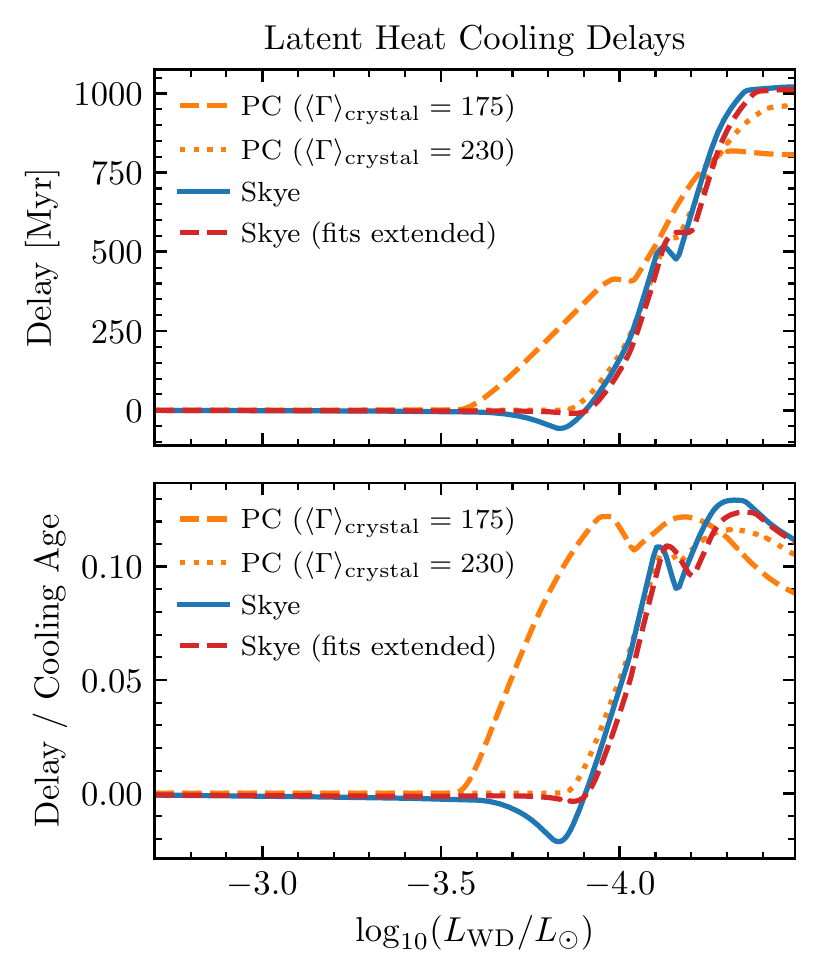}
  \caption{Comparison of latent heat cooling delays for \skye\ and PC
    with crystallization occurring at two different values of
    $\langle \Gamma \rangle$. All delays are relative to a model run
    using the PC EOS with $\langle \Gamma \rangle_{\rm crystal} = 230$
    and no latent heat release.}
  \label{fig:delay}
\end{figure}

Figure~\ref{fig:delay} shows the cooling delay introduced into WD
models by latent heat from crystallization in models run with each of the PC EOS and \skye.
For \skye\ we performed two sets of calculations, one with the default extrapolation settings and another `fits extended' calculation where we used $\Gamma_{\rm min}^{\rm solid}=100$ and $\Gamma_{\rm max}^{\rm liquid} = 300$.

In general the \skye\ models agree well with the PC model run with
crystallization occurring at
$\langle \Gamma \rangle = 230$, which represents the previous state of
the art for WD cooling in MESA. Before crystallization begins around
$\log_{10}(L/L_\odot) = -3.8$, the lower panel of
Figure~\ref{fig:delay} also shows that the \skye\ WD models agree with
the overall cooling age of the PC model to better than 1\%.

The \skye\ models also agree well with each other despite the `fits extended' version applying the free energy fits over a wider range of temperatures.
The reason for this is that  \skye\ is thermodynamically consistent, so the overall cooling delay produced by the phase transition is insensitive to the choice of $\Gamma_{\rm max}^{\rm liquid}$ and $\Gamma_{\rm min}^{\rm solid}$.
To see this note that the entropy deep in the liquid phase (all $\Gamma_j < \Gamma_{\rm max}^{\rm liquid}$) is independent of the extrapolation process, and likewise for the entropy deep in the solid phase (all $\Gamma_j > \Gamma_{\rm min}^{\rm solid}$).
Hence, if the temperature varies little across the transition and extrapolation window then $\int T \partial s/\partial T dT$, counting the $\epsilon_{\rm latent}$ term, is nearly independent of the extrapolation limits.

\begin{figure}
  \centering
  \includegraphics[width=0.45\textwidth]{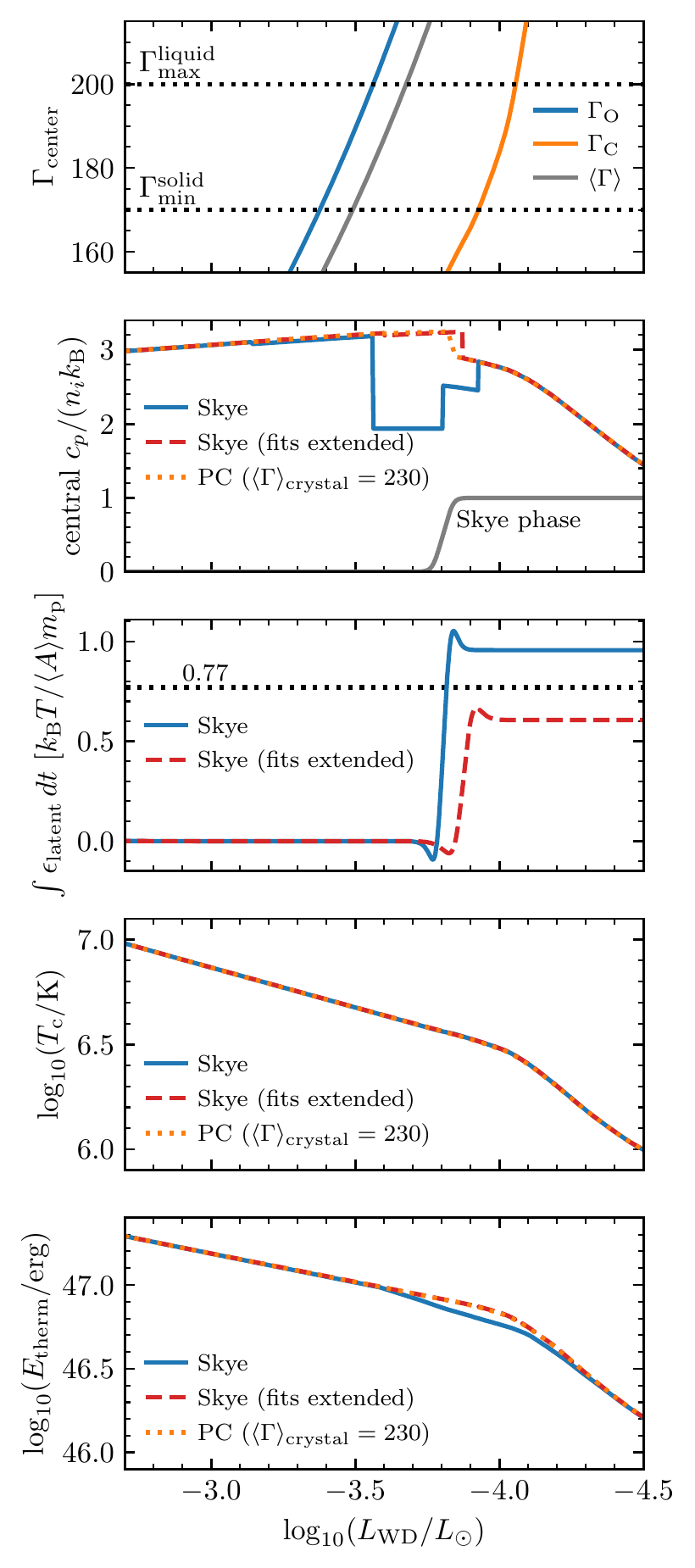}
  \caption{Core thermodynamic properties as a function of luminosity
    for WD cooling models running on Skye and PC. In the second panel \skye\ phase refers to the quantity $\phi$ from equation~\eqref{eq:phi}.
  }
  \label{fig:LTc}
\end{figure}

Figure~\ref{fig:LTc} gives a comparison of the interior properties of
the WD cooling models from Skye and PC with crystallization at $\langle \Gamma \rangle = 230$.
As expected, the $L$--$T_{\rm c}$ relation agrees very well between Skye
and PC models, reflecting the similar input physics underlying these
two EOSs. Similarly, the total WD thermal content, defined as
$E_{\rm therm} \equiv \int c_p T \, dm$, agrees very well between the
two models.

The heat capacity $c_p$ in Figure~\ref{fig:LTc} shows some
disagreement in the region near the phase transition from liquid to solid.
The notch-like behavior in the \skye\ $c_p$ is a result of our thermodynamic extrapolation prescription (Section~\ref{sec:extrapolate}).
This is because the one-component plasma contribution to $\partial s/\partial T$ vanishes when we extrapolate, so the contribution to $c_v$ vanishes:
\begin{align}
	c_v &=  \left.\frac{\partial e}{\partial T}\right|_{\rho}\nonumber\\
	&= \left.\frac{\partial (F+Ts)}{\partial T}\right|_{\rho}\nonumber\\
	&= \left.\frac{\partial F}{\partial T}\right|_{\rho} + s + T\left.\frac{\partial s}{\partial T}\right|_{\rho}\nonumber\\
	&= T\left.\frac{\partial s}{\partial T}\right|_{\rho}.
\end{align}
Therefore, for any species at a $\Gamma_j$ where its free energy is being extrapolated, its OCP contribution to $c_v$ vanishes.
Because $c_p \sim c_v$, this causes a drop in $c_p$ as well.
Reading from left to right in the $c_p$ panel of Figure~\ref{fig:LTc}, the core begins in the liquid phase and initially no extrapolation is needed for the liquid phase free energy because $\Gamma_j < \Gamma_{\rm max}^{\rm liquid}$ for all species.
As the core cools, the $\Gamma_j$ rise.
The heat capacity falls sharply when $\Gamma_{^{16}\rm O}$ reaches $\Gamma_{\rm max}^{\rm liquid}$ because past that point we extrapolate the OCP free energy of $^{16}\rm O$.
The core continues cooling and then crystallizes at $\log L/L_\odot = -3.8$.
At this point the heat capacity is determined by the solid phase free energy.
Because  $\Gamma_{^{16}\rm O} > \Gamma_{\rm min}^{\rm solid}$, the OCP free energy of $^{16}\rm O$ is no longer extrapolated, but $\Gamma_{^{12}\rm C} < \Gamma_{\rm min}^{\rm solid}$ so the free energy of $^{12}\rm C$ is now extrapolated in the solid phase.
Finally, once $\log L/L_\odot = -3.9$, $\Gamma_{^{12}\rm C} > \Gamma_{\rm min}^{\rm solid}$ so we stop extrapolating the $^{12}\rm C$ free energy, causing a jump in $c_p$.
At this stage no species are extrapolated, and the heat capacity remains smooth for the rest of the run.

As before we note that because \skye\ is thermodynamically consistent the overall cooling delay is insensitive to the choice of limits for thermodynamic extrapolation and hence to these features in $c_p$.
So for instance in Figure~\ref{fig:LTc} extrapolation reduces $c_p$ near the phase transition relative to the `fits extended' version of \skye.
The third panel of Figure~\ref{fig:LTc} shows the total latent heat released in the core in terms of the thermal energy per ion at the temperature of crystallization, and we see that this is decreased for the `fits extended' version.
Thus the \emph{decrease} in $c_p$ is offset in the overall cooling calculating by an \emph{increase} in $\epsilon_{\rm latent}$, resulting in the regular and `fits extended' versions of \skye\ showing very similar cooling curves in Figure~\ref{fig:delay}.

In both the regular and `fits extended' versions of \skye\ we see that the overall magnitude of the latent heat is similar to the value of
$0.77 k_{\rm B}T/\langle A \rangle m_{\rm p}$ calculated by
\cite{Salaris2000}, which has often been adopted in recent studies of
WD cooling using other stellar evolution codes (e.g.,
\citealt{Camisassa2019}).
It is likewise similar to the results of~\citet{2013A&A...550A..43P}, who obtained an improved value of $0.75 k_\mathrm{B} T / A m_\mathrm{p}$ in the case of the one component plasma with the `rigid' electron backgroung and showed that the allowance for electron polarization/screening can lead to deviations of up to a factor of two from this value.

In our testing these sharp features in $c_p$ have not caused any convergence problems in MESA.
However, if this behavior is undesirable, $\Gamma_{\rm min}^{\rm solid}$ can be lowered and $\Gamma_{\rm max}^{\rm liquid}$ can be raised to ensure that, for any given composition, extrapolation is only used for the liquid phase when the system is solid, and vice versa, with the caveat that this risks using fitting formulas beyond the region in which they are known to be accurate.
This is what is shown in the `fits extended' curves in Figures~\ref{fig:delay} and~\ref{fig:LTc}, where we used $\Gamma_{\rm min}^{\rm solid}=100$ and $\Gamma_{\rm max}^{\rm liquid} = 300$.
Our hope is that future work on multi-component plasmas will provide a way to capture the behavior of, e.g., low-$\Gamma$ carbon in a multi-component solid.
This could take the form of e.g., fits for the two-component plasma free energy at the phase transition as a function of the charge ratio between the two species.

\begin{figure}
  \centering
  \includegraphics[width=0.45\textwidth]{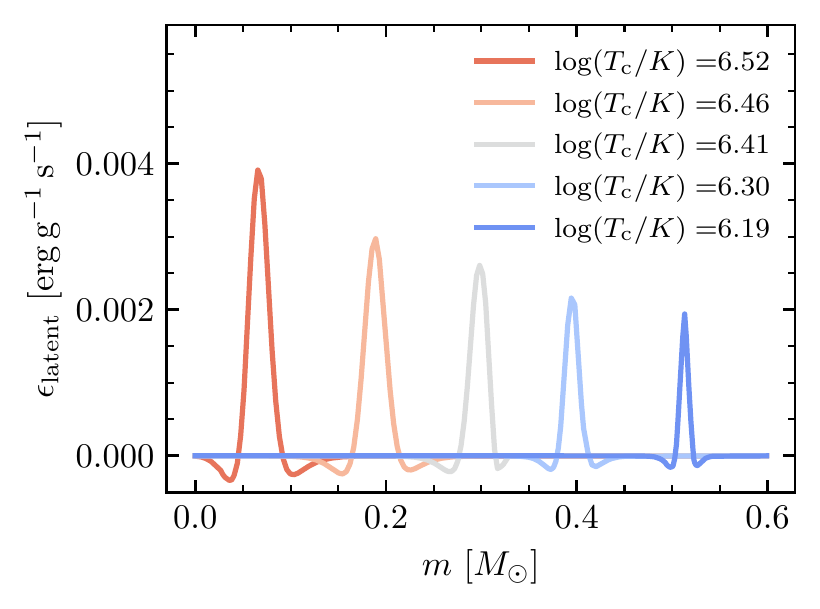}
  \caption{Evolution of the latent heating term from \skye\ as the WD
    model cools and the crystallization front moves from the center
    toward the surface.}
  \label{fig:eps_latent}
\end{figure}

\begin{figure}
  \centering
  \includegraphics[width=0.45\textwidth]{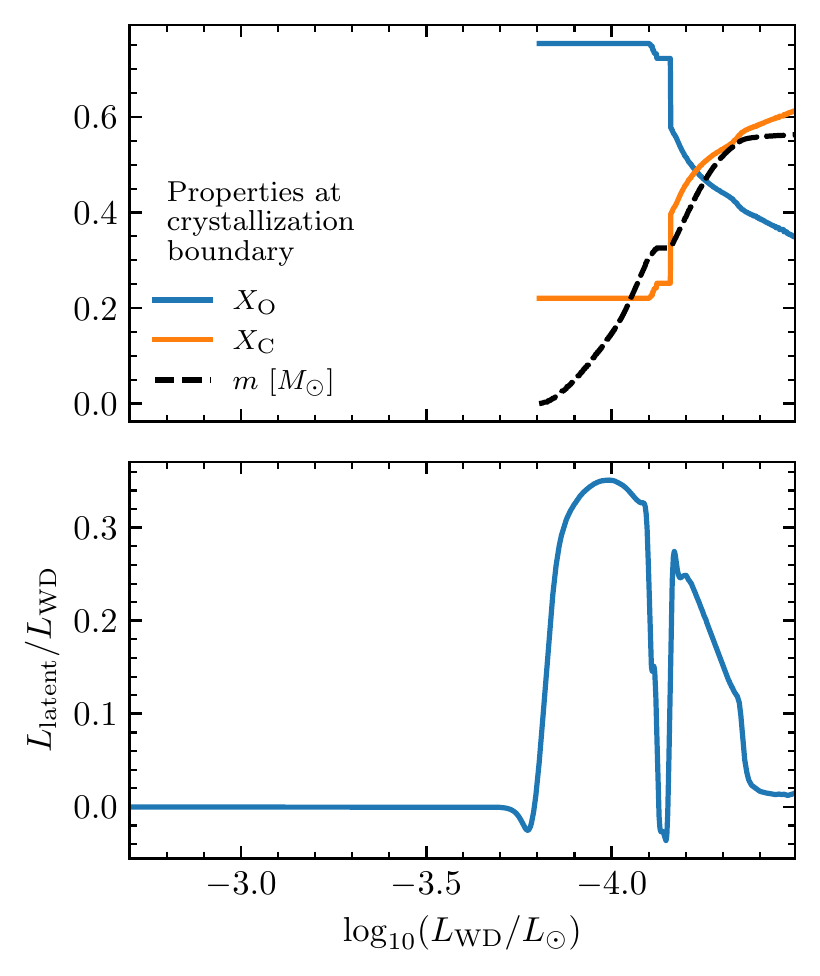}
  \caption{{\it Upper panel:} Curves showing the mass coordinate and
    composition of material at the crystallization boundary as a
    function of WD luminosity.
    {\it Lower panel:} Total luminosity from latent heating as a fraction of the
    WD luminosity, where $L_{\rm latent} \equiv \int \epsilon_{\rm latent} \, dm$.}
  \label{fig:l_latent}
\end{figure}

Figures~\ref{fig:eps_latent} and~\ref{fig:l_latent} show more details
about the latent heating term from Skye in our WD cooling model.
Figure~\ref{fig:eps_latent} shows how the blurred phase
transition distributes the latent heat in the WD interior as the
crystallization front moves outward while the WD cools. Integrating
these heating profiles over the entire WD gives a total latent heating
luminosity $L_{\rm latent}$, which is shown in Figure~\ref{fig:l_latent}.
The upper panel of that figure also shows the composition and mass
coordinate location of the crystallization boundary (defined as the
location where Skye phase = 0.5). We note that as the crystallization
front moves outward, there is a brief pause in crystallization and the
latent heating goes to zero when the front reaches a location where
the core composition becomes more carbon-rich. This location
corresponds to the outer edge of the former convective He-burning
core at the end of central He-burning, where C/O layers exterior to
this point were produced by subsequent He shell burning and therefore
have a different C/O composition than the interior homogeneous core.
This relatively carbon-rich layer has a lower crystallization
temperature than the adjacent C/O core interior to it, and so the core
temperature must cool further before crystallization resumes and the
latent heat returns.

\begin{figure}
  \centering
  \includegraphics[width=0.45\textwidth]{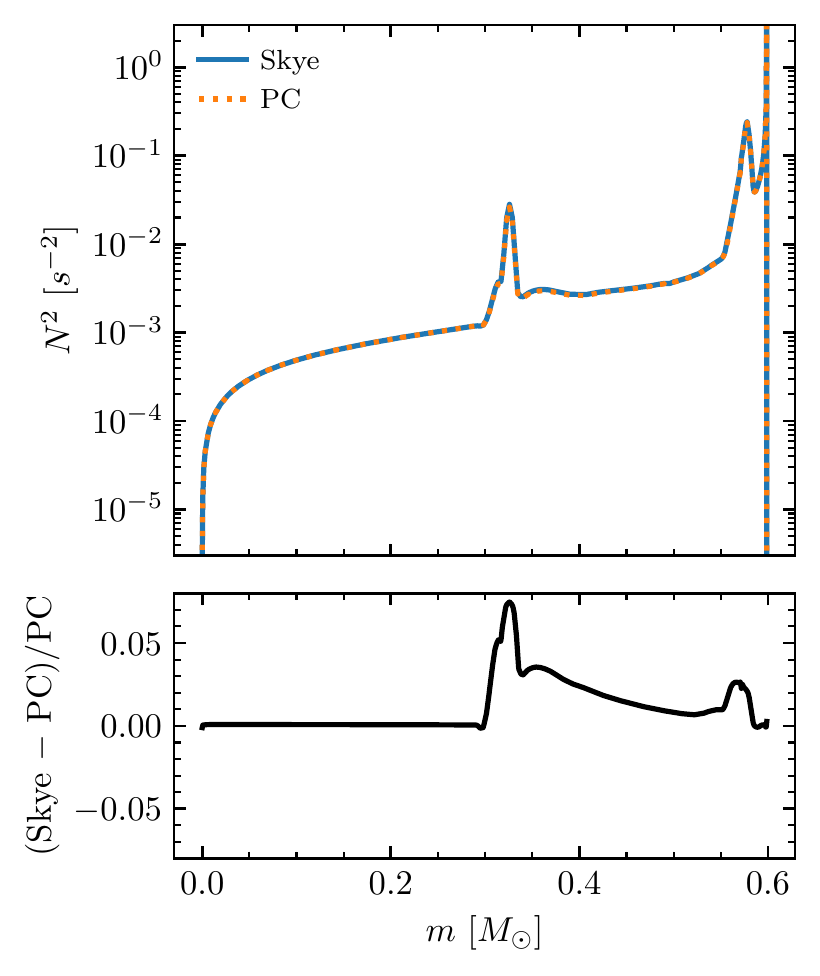}
  \caption{{\it Upper panel:} The \brvs\ frequency is shown as a function of mass coordinate for the \skye\ and PC WD models at a time when $\log T_{\rm c}/\mathrm{K} = 7.05$ and $T_{\rm eff} = 11,800\,\mathrm{K}$.  {\it Lower panel:} The relative difference between the two models is shown as a function of mass coordinate.}
  \label{fig:BV}
\end{figure}

Finally, Figure~\ref{fig:BV} shows the profile of the \brvs\ frequency for both the \skye\ and PC WD models, as well as the relative difference between the two.
The differences are generally of order a few percent.
For $m > 0.3 M_\odot$, there are differences in the composition gradient region.
These arise because \skye\ treats the density $\rho$ as the baryonic mass density whereas PC treats it as the physical mass density.
Either choice is valid, but neither is fully consistent with how \code{MESA} computes either the \brvs\ frequency or hydrostatic equilibrium, and these inconsistencies produce the differences we see for $m > 0.3 M_\odot$.

\section{Execution Efficiency}
\label{sec:speed}

\skye\ is designed to be fast enough to evaluate at runtime in stellar evolution calculations.
We benchmarked \skye, HELM, and PC on a single core of an Intel Core i9 (I9-9980HK) CPU running at 2.4GHz.
For this test PC was modified to use CR-LIBM for mathematical operations to ensure bit-for-bit identical results across platforms just like \skye\ and HELM.

We evaluated each EOS on a log-spaced grid in $\rho$ spanning $10^{-10}-10^{10}\,\mathrm{g\,cm^{-3}}$ with 600 points and in $T$ spanning $10^3-10^{10}\,\mathrm{K}$, with 500 points.
We require each EOS to return all of the quantities listed in Section~\ref{sec:thermo} except for the \skye-specific ones, as well as the partial derivatives of each of those quantities with respect to $\rho$ and $T$.
Because PC does not natively provide those derivatives, we use three calls of PC per point and then extract the additional derivatives with finite differences.

Averaged over all points in our grid, \skye\ takes $17\mu\mathrm{s}$ per call, PC takes $9\mu\mathrm{s}$ per call, and HELM takes $6\mu\mathrm{s}$ per call, where again we evaluate PC three times per call to produce the additional derivatives required by stellar evolution software instruments such as MESA.

As a second benchmark, we tracked the time spent in the MESA EOS module during the white dwarf cooling study from Section~\ref{sec:cooling}.
The EOS accounted for 10.5 per-cent of total run time when using PC, and 13.9 per-cent of total run time when using \skye.
This understates the difference between the two slightly because some of the time the stellar model is at a temperature and density where neither PC nor \skye\ are used, but shows that the runtime difference is minimal not only on a grid but also in practice in stellar evolution calculations.

\skye\ and PC have similar performance for several reasons:
\begin{enumerate}
	\item The physics that enters these equations of state is similar.
	\item Our automatic differentiation type is heavily optimized, and in many cases produces performance similar to hand-coded derivatives.
	\item The additional cost of determining higher-order derivatives with automatic differentiation happens to be very similar to the overhead of calling PC three times to obtain the same derivatives with finite differences.
	\item While \skye\ has to compute the non-ideal free energy twice to obtain phase information, this extra cost relative to PC is offset by the fact that \skye\ uses free energy tables for the ideal electron-positron contribution while PC computes this with more expensive fitting formulas.
\end{enumerate}
We determined (3) by producing a modified version of PC which produces higher-order derivatives using automatic differentiation rather than finite differences and found its performance to be similar to the unmodified PC.

HELM is much faster than either \skye\ or PC for three main reasons.
First, HELM uses an average composition characterized by the mean molecular weight and mean charge, 
rather than directly using the full composition vector $\{y_j\}$.
Second, the computationally expensive parts of HELM (a root-find for the electron chemical potential, 
high precision Fermi-Diac integrals, and nearly all operations involving division, exponentials, and power functions) are tabulated 
on a logically rectalinear array. Each call to HELM then consists of hash table lookups followed 
by calls to fast polynomial interpolation functions.
Third, thermodynamic information for neighboring points are located next to each other in physical memory. 
Ordered sweeps, such as from the surface of a stellar model to the center, 
will usually access data already loaded into the processor cache rather than having to access 
data from the slower main memory. This reduction in the time required to access
information from memory boosts the execution efficiency.

\section{Availability} \label{sec:avail}

\skye\ is distributed as part of the \code{eos} module of the \code{MESA} stellar evolution software instrument.
It is also available as a standalone package from \url{https://github.com/adamjermyn/Skye}, and the version used here is available from~\citet{adam_s_jermyn_2021_4641111}.
Compilation is supported on the GNU Fortran compiler version 10.2.0.

\section{Future Work}\label{sec:conc}

Because \skye\ is a framework for developing new EOS physics we expect future work to bring several key improvements.
First, and most pressing, is handling of partial ionization and neutral matter.
With that \skye\ could be used across the entire range of densities and temperatures which arise in stellar evolution calculations.
This could be done in a Debye-Huckle-Thomas-Fermi formalism~\citep{PhysRev.111.1460} or other approaches in the physical picture~\citep{Rogers2002}, or else via free energy minimization~\citep{Irwin2004} in the chemical picture~\citep{Saumon1995}.
The key constraint in each of these approaches is that \skye\ needs to remain fast enough to use in practical stellar evolution calculations.
Our hope is that the flexibility afforded to \skye\ by its automatic differentiation machinery will allow us to rapidly prototype and test these various possibilities.

Along similar lines, \skye\ could be made to support phase separation by minimizing the free energy with respect to the compositions of the liquid and solid phases.
The major bottleneck to supporting this is the current lack of Fortran compiler support for parameterized derived types.
Once this compiler challenge is resolved, phase separation physics  should not be difficult to implement.

More broadly, we make \skye\ openly available with the hope that it will grow into a community resource
to use automatic differentiation to explore analytic free energy terms that captures improvements in
existing physics and development of new or not yet considered physics.

\begin{acknowledgments}

The Flatiron Institute is supported by the Simons Foundation.
We thank Lars Bildsten for conversations and thoughts on equations of state,
and Simon Blouin for providing the phase curve results from \citet{Blouin2020} in machine-readable form.
We are grateful to Gilles Chabrier for helpful comments on this manuscript and for his contribution to the PC EOS, and some of the routines in \skye\ are based on routines from PC.
ASJ is grateful to Dan Foreman-Mackey for early conversations on automatic differentiation.
FXT is indebted to Werner D\"{a}ppen, and Doug Swesty for sharing their knowledge and source codes on equations of state over the years.
ASJ and EB thank the Gordon and Betty Moore Foundation (Grant GBMF7392) and the National Science Foundation (Grant No. NSF PHY-1748958) for supporting this work.
The MESA project is supported by the National Science Foundation (NSF) under the Software Infrastructure for Sustained Innovation
program grants (ACI-1663684, ACI-1663688, ACI-1663696).
This research was also supported by the NSF under grant PHY-1430152 for the Physics Frontier Center
``Joint Institute for Nuclear Astrophysics - Center for the Evolution of the Elements'' (JINA-CEE).
The work of AYP was partially supported by The Ministry of Science and Higher Education of the Russian Federation
(Agreement with Joint Institute for High Temperatures RAS No.\,075-15-2020-785).

\end{acknowledgments}

\software{
\texttt{Skye} \url{https://github.com/adamjermyn/Skye},
\texttt{PC} \citep[][\url{http://www.ioffe.ru/astro/EIP/index.html}]{CP1998,PhysRevE.62.8554,PhysRevE.79.016411,Potekhin2010},
\texttt{HELM} \citep[][\url{http://cococubed.asu.edu/code_pages/eos.shtml}]{Timmes2000},
\texttt{MESA} \citep[][\url{http://mesa.sourceforge.net}]{Paxton2011,Paxton2013,Paxton2015,Paxton2018,Paxton2019},
\texttt{MESASDK} 20190830 \citep{mesasdk_linux,mesasdk_macos},
\texttt{CR-LIBM} \citep[][\url{http://www.ens-lyon.fr/LIP/AriC/ware}]{CR-LIBM}, 
\texttt{matplotlib} \citep{hunter_2007_aa}, 
\texttt{NumPy} \citep{der_walt_2011_aa}, and
\texttt{SymPy} \citep{10.7717/peerj-cs.103}.
         }

\clearpage

\appendix

\section{MESA} \label{appen:mesa}

\begin{figure}
	\centering
	\includegraphics[width=0.45\textwidth]{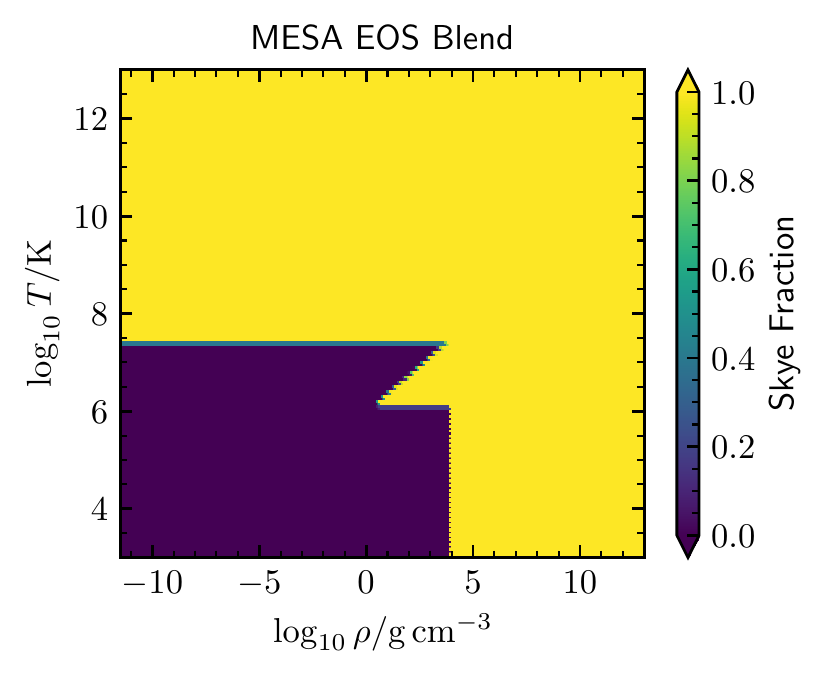}
	\caption{The fraction of \skye\ used in the MESA EOS is shown as a function of density and temperature.}
	\label{fig:blend}
\end{figure}

Our calculations of stellar structure and evolution were performed with commit \texttt{21fd6fa} of the MESA software instrument, based upon the recent release r15140.
We patched this commit to use the version of PC which ships with MESA revision 12778 because that is more similar to the original PC EOS.
MESA uses a blend of \skye, OPAL \citep{Rogers2002}, SCVH
\citep{Saumon1995}, FreeEOS \citep{Irwin2004}, and HELM~\citet{Timmes2000}.
The blend uses \skye\ in most of the region where $T > 10^{6.2}\,\mathrm{K}$ or $\rho > 10^{4}\,\mathrm{g\,cm^{-3}}$, though the precise shape of the blend between this EOS and the others is more complicated than a simple cutoff (see Figure~\ref{fig:blend}), and was determined to minimize the the difference in energy between equations of state across the blend.

Radiative opacities are primarily from OPAL \citep{Iglesias1993,
Iglesias1996}, with low-temperature data from \citet{Ferguson2005}
and the high-temperature, Compton-scattering dominated regime by
\citet{Poutanen2017}.  Electron conduction opacities are from
\citet{Cassisi2007}.

Nuclear reaction rates are a combination of rates from
NACRE \citep{Angulo1999}, JINA REACLIB \citep{Cyburt2010}, plus
additional tabulated weak reaction rates \citet{Fuller1985, Oda1994,
Langanke2000}. Screening is included via the prescription of \citet{Chugunov2007}.  Thermal
neutrino loss rates are from \citet{Itoh1996}.

\section{EOS Comparisons} \label{appen:helm}

For standalone EOS comparisons we use the version of PC which ships with MESA revision 12778, which notably smooths thermodynamic quantities across the phase transition.
This was a modification made for numerical reasons in MESA, but should not substantially affect the substance of our comparisons.
We disable Coulomb corrections in HELM and enforce full ionization across the $\rho-T$ plane.
We use the tabulated free energy for all HELM quantities, including $\partial p/\partial \rho|_T$ and $\partial^2 p/\partial\rho^2|_T$, rather than the auxiliary tables which provide these separately.
High quality numerical derivatives were determined using the $\mathrm{dfridr}$ option in the $\mathrm{eos\_plotter}$ routine in MESA.

\section{Data Availability}\label{appen:data}

The data and related scripts used in this work are available at~\citet{adam_s_jermyn_2021_4639793}.

\section{Phase Transitions and Quantum Corrections}
\label{sec:quantum}

\begin{figure}
	\centering
	\includegraphics[width=0.45\textwidth]{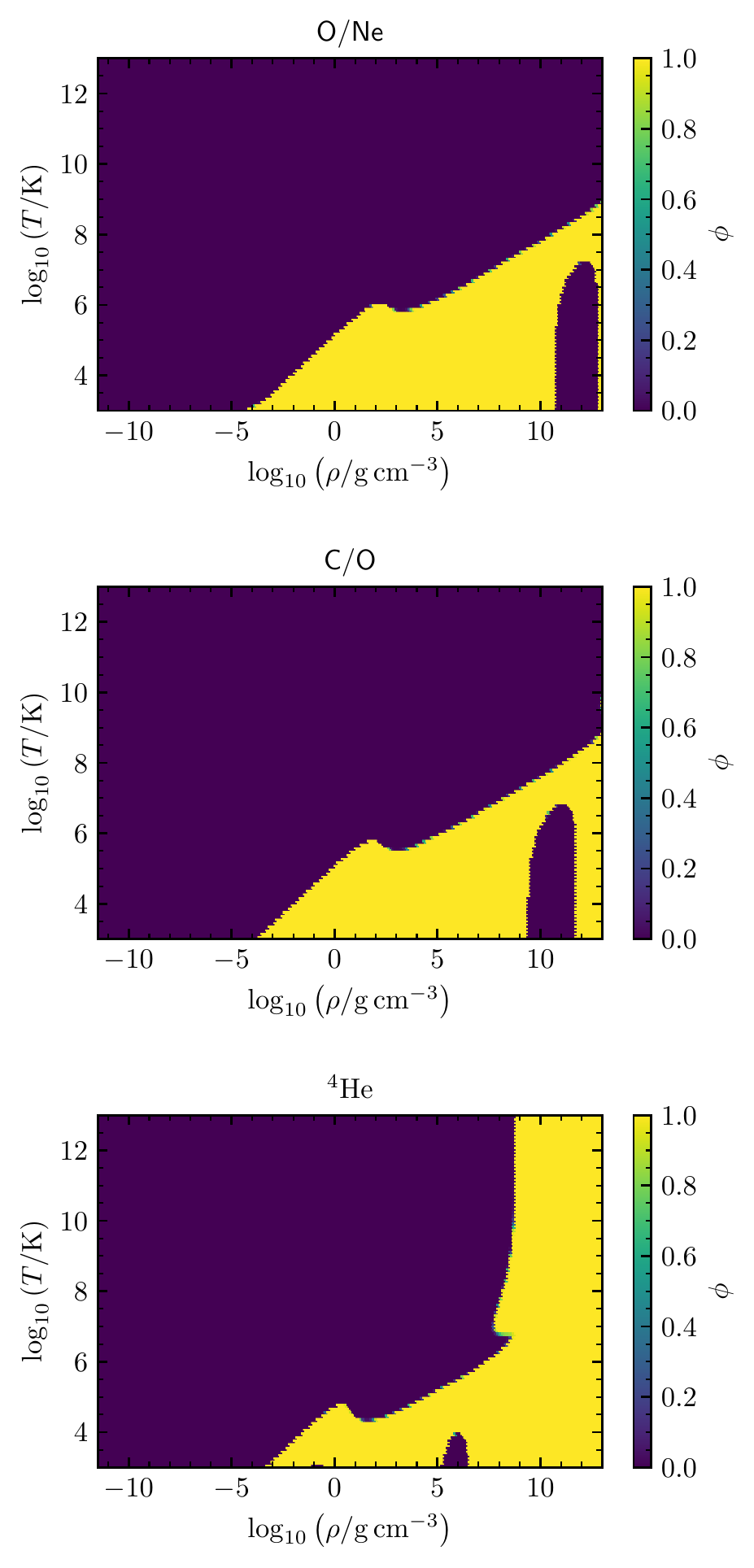}
	\caption{The phase $\phi$ is shown as a function of temperature and density for (upper) an equal-mass mixture of $^{16}\rm O$ and $^{20}\rm Ne$, (middle) an equal-mass mixture of $^{12}\rm C$ and $^{16}\rm O$, and (lower) pure $^{4}\rm He$.}
	\label{fig:3_phase}
\end{figure}

Figure~\ref{fig:3_phase} shows the \skye\ phase $\phi$ as a function of $\rho$ and $T$ for three different compositions.  At high temperatures and low densities the system is a liquid, and it crystallizes in the opposite limit.  This standard OCP-like phase transition that occurs at approximately constant $\langle \Gamma \rangle$ is discussed in the main text.  However, Figure~\ref{fig:3_phase} displays additional structure in the phase, which we determined to be primarily related to the quantum correction terms in the free energy.  These features likely reflect limitations in the assumed prescriptions.

At high densities for the lightest elements ($\mathrm{H}$ and $\mathrm{He}$), quantum corrections dominate and favor the solid phase up to high temperatures.  While a self-consistent consequence of the adopted inputs, we suspect this feature is spurious.  However, as $^{4}\mathrm{He}$ and $^{1}\mathrm{H}$ are likely to have fused into heavier elements long before reaching these densities in typical astrophysical applications, 
we have done nothing to suppress this solidification in \skye.

\begin{figure}
	\centering
	\includegraphics[width=0.45\textwidth]{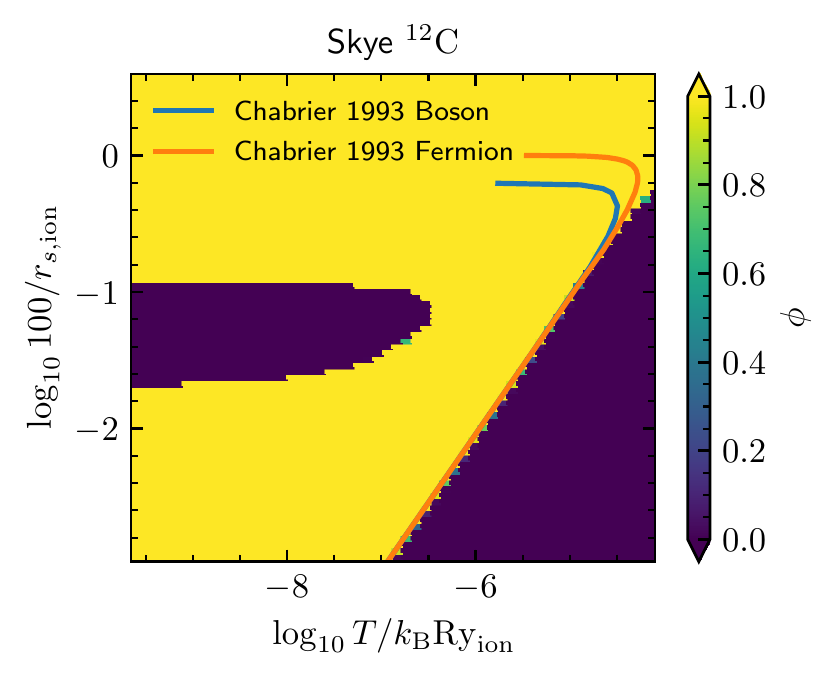}
	\caption{The phase $\phi$ is shown as a function of temperature in units of $\mathrm{Ry}_{\rm ion}$ and reciprocal ion spacing measured by $100/R_{S,\rm ion}$. The melt lines of~\citet{1993ApJ...414..695C} for bosons (blue) and fermions (orange) are over-plotted for comparison. This calculation was done for pure $^{12}\mathrm{C}$, but the choice of units means the results are universal for any pure ionic system.}
	\label{fig:C_phase}
\end{figure}

At high densities and at low temperatures, quantum corrections dominate and cause the system to melt.
This occurs at lower densities and temperatures for lower-mass lower-charge species: $10^{10}\,\mathrm{g\,cm^{-3}}$ for O/Ne, $10^{8}\,\mathrm{g\,cm^{-3}}$ for C/O, and $10^4\,\mathrm{g\,cm^{-3}}$ for $^{4}\mathrm{He}$).
A similar effect has been seen in Monte Carlo calculations and analytic calculations~\citep{1993ApJ...414..695C,PhysRevB.18.3126,PhysRevLett.76.4572}.
In those studies the Lindemann criterion was used to compute the quantum melt line, but the result has a rather different topology from the phase  boundary we see (Figure~\ref{fig:C_phase}).
In particular we see the quantum melt only for a finite density range, whereas they predict it for all densities above a cutoff.
The latter is more in line with our understanding of the physics of quantum melting, namely that it is driven by the zero-point energy of ions and so should only increase with increasing density.
We therefore suspect that the topology of this melt region reflects limitations in our prescriptions for the OCP quantum corrections.

Moreover the temperature and density scale involved is rather different from Lindemann criterion calculations~\citep{1993ApJ...414..695C,PhysRevB.18.3126,PhysRevLett.76.4572}, though interestingly the \emph{scaling} of these scales matches those from the Lindemann criterion.
The melt line is predicted to peak around $k_{\rm B} T \approx 6\times 10^{-5}\,\mathrm{Ry}_{j}$, where
\begin{align}
	\mathrm{Ry}_{j} = (Z_{j} e)^4 m_{j} / 2 \hbar^2
\end{align}
is the ionic Rydberg.
Instead we see a peak near $6\times 10^{-6}\,\mathrm{Ry}_{\rm ion}$.
Likewise the melt line is predicted to peak in temperature when the dimensionless ion sphere radius
\begin{align}
	r_{s,j} \equiv \left(\frac{3 m_j}{4\pi \rho}\right)^{1/3}  \frac{m_j (Z_j e)^2}{\hbar^2}
\end{align}
is of order $200$, and we see the peak around $1200$.

Overall the disagreement between \skye\ and calculations based on the Lindemann criterion suggests caution in interpreting these results.
This disagreement may be caused by our use of the fit by~\citet{2019MNRAS.490.5839B} beyond its range of validity, which is confined within the dark blue triangle at the lower right corner of Figure~\ref{fig:C_phase}.
These results are, however, a completely self-consistent consequence of the input physics
so we have not done anything to impede quantum melting in \skye.

\bibliography{refs}
\bibliographystyle{aasjournal}

\end{document}